\documentclass{aa}
\usepackage{graphicx}    
\usepackage{txfonts}     
\usepackage{subcaption} 
\usepackage{lscape}     
\usepackage{placeins}
\usepackage{placeins}   
\usepackage{afterpage} 
\usepackage{multirow}
\usepackage{xcolor}
\usepackage[
    colorlinks=true,
    linkcolor=red,
    citecolor=blue,
    urlcolor=blue
]{hyperref}

\begin{document}
\authorrunning {Zullo et al.}
\titlerunning {JWST observations of Terzan 5}

\title{The multi-age stellar  populations of Terzan 5 as revealed by JWST\thanks{Based on observations made with the NASA/ESA/CSA James Webb Space Telescope under program GO5502 (PI:Ferraro)}}

\author{G.~Zullo\inst{1,2}, C. Pallanca\inst{1,2}, F.R. Ferraro\inst{1,2}, B. Lanzoni\inst{1,2}, L. Origlia\inst{2}, D. Massari\inst{2}, E. Dalessandro\inst{2}, C. Fanelli\inst{2}, M. Cadelano\inst{1,2}, E. Vesperini\inst{3}, C. Crociati\inst{4}, R.M. Rich\inst{5}, E. Valenti\inst{6,7} }

\institute{
  Dipartimento di Fisica e Astronomia, Università degli Studi di Bologna, 
  Via Piero Gobetti 93/2, I-40129 Bologna, Italy
  \and
  INAF – Osservatorio di Astrofisica e Scienza dello Spazio di Bologna (OAS), 
  Via Piero Gobetti 93/3, I-40129 Bologna, Italy
   \and
Dept. of Astronomy, Indiana University, Bloomington, IN 47401, USA
\and
Institute for Astronomy, University of Edinburgh, Royal Observatory, Blackford Hill, Edinburgh, EH93HJ, UK
\and
Department of Physics and Astronomy, UCLA, 430 Portola Plaza, Box 951547, Los Angeles, CA 90095-1547, USA
\and European Southern Observatory, Karl-Schwarzschild-Strasse 2, 85748 Garching bei Munchen, Germany
\and
Excellence Cluster ORIGINS, Boltzmann-Strasse 2, D-85748 Garching Bei Munchen, Germany\\}
\date{March 30, 2026}

\abstract{
The James Webb Space Telescope provides an exciting opportunity to investigate stellar systems located in heavily obscured regions like the Galactic bulge. Possibly, the most enigmatic among them is Terzan 5: long classified as a globular cluster, it is now known to host distinct stellar populations with different iron abundances (ranging  approximately from [Fe/H]=$-0.8$ to  [Fe/H]=$+0.3$ dex).  Indeed the chemical and structural properties collected so far suggest that it is the remnant of one of the primordial clumps that contributed to the early assembly of the bulge, a so-called "Bulge Fossil Fragment". Here we present a new photometric analysis of Terzan 5 based on JWST/NIRCam observations in the F115W and F200W filters, as well as archival HST/ACS optical (F606W and F814W) data. The dataset overcomes the severe and spatially variable extinction along the line of sight and yields the deepest color–magnitude diagram ever obtained for Terzan 5. Proper motion selections and high-resolution differential reddening corrections allow us to isolate bona fide cluster members and to provide an unprecedented view of the main-sequence turn-off region. 
We clearly identify two main components and determine their respective ages: the old, sub-solar component has an age of 12.5 $\pm$ 0.5 Gyr,  while the super-solar component is significantly younger with an age of 4.7 $\pm$ 0.5 Gyr. Interestingly, we also find hints of an even younger main sequence turn-off and sub-giant branch, consistent with the presence of a further stellar component with an age of 3.8 $\pm$ 0.5 Gyr. There is also evidence of a blue plume populated by stars as bright as $m_{\rm F115W}\sim 17.4$, suggesting a prolonged period of star formation extending up to 2.5 Gyr ago.}
\keywords{techniques: photometric --
          Galaxy: bulge --
          Galaxy: stellar content --
          globular clusters: individual: Terzan 5}

\maketitle
\nolinenumbers

\section{Introduction}
\begin{figure}[t]
  \centering
  \includegraphics[width=0.5\textwidth]{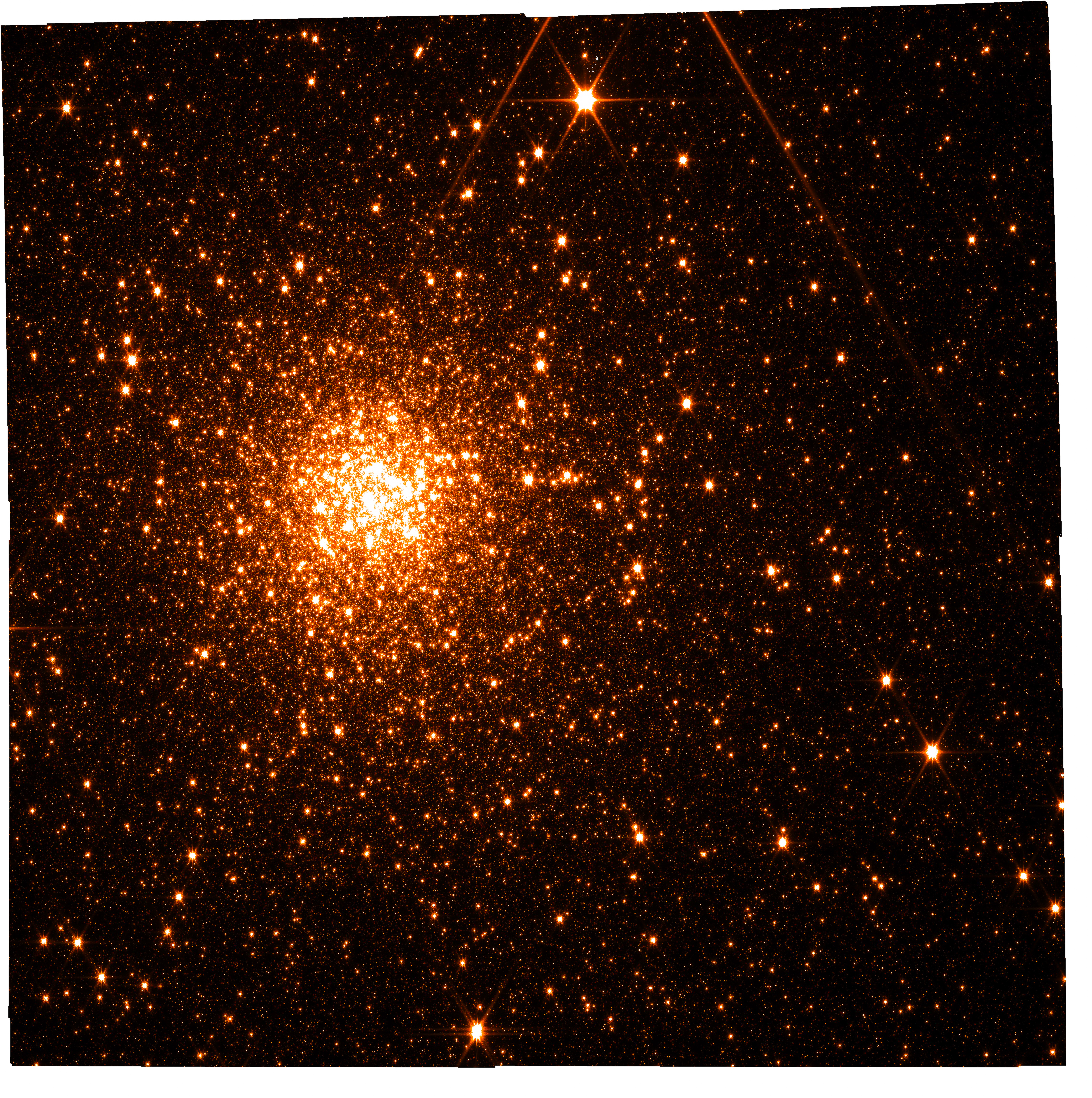}
  \caption{JWST/NIRCam mosaic of Terzan 5 obtained with module B in the F115W filter (composite of all dithered exposures from SW detectors onto a common grid).}
  \label{fig:Terzan 5nircam}
\end{figure}
Terzan 5 stands out as one of the most enigmatic stellar systems in the Galactic bulge.   Located at approximately 2 kpc from the Galactic center
and obscured by severe and spatially variable extinction with average color excess $E(B-V)=2.38 \pm 0.55 $ (\citealt{valenti2010,massari2012}), it has long been both alluring and challenging to observe. 
Among its notable characteristics are its unusually high mass—with published stellar-mass estimates of $ (2 \pm 1)\times10^6M_\odot$ (\citealt{Lanzoni2010}) and $(1.09 \pm 0.08)\times10^6M_\odot$ (\citealt{baumgardt2021})—and the presence of the largest known population of confirmed millisecond pulsars in any Galactic globular cluster (GC), with the most recent count reaching 49 objects \citep{ransom, Cadelano_2018, Padmanabh2024}. Additionally, Terzan 5 displays the highest collision rate among Galactic GCs \citep{Lanzoni2010}.
However, its most remarkable feature has emerged from extensive photometric and spectroscopic studies conducted over the last sixteen years, following the identification of 
at least two main stellar sub-populations 
characterized by distinct iron abundances \citep{ferraro2009, Massari2014, origlia2011}. Initially suggested by the presence of two clearly separated horizontal branches in the color-magnitude diagram 
\citep[CMD;][]{ferraro2009}, this was later confirmed by subsequent spectroscopic studies \citep{Massari2014, origlia2011,  origlia2013, origlia2025}, which revealed two major [Fe/H] peaks at $-0.25 \pm 0.07 $ and $+0.27 \pm 0.04$ dex, alongside a third, less prominent component at [Fe/H] $= -0.79 \pm 0.04$ dex. Such a broad, multi-peaked 
%high-metallicity
[Fe/H] distribution is unprecedented among Galactic GCs. Moreover, Terzan 5 shows no sign of an Al-O anti-correlation \citep{origlia2011} as typically observed in GCs. The main metal-poor population was also found to be $\alpha$-enhanced ([$\alpha$/Fe] $\simeq +0.34$ dex), whereas the metal-rich component exhibits nearly solar abundances \citep[${\rm [\alpha/Fe]} \simeq 0.03$ dex;][]{origlia2011,origlia2013,origlia2019,origlia2025}. The position of the knee in the [$\alpha$/Fe]–[Fe/H] relation implies an early star-formation efficiency significantly higher than that of the Galactic halo \citep{helmi2020},
the Galactic disk \citep{Haywood2018}, and dwarf galaxies \citep{Tolstoy}. On the contrary, 
the chemical abundances measured from
the entire high-resolution spectroscopic dataset 
acquired so far for Terzan 5 fully match the chemical pattern of the Galactic bulge itself \citep{origlia2025}. This  
strongly supports a tight evolutionary link that 
points to a bulge in-situ formation for Terzan 5.
The first age estimates for Terzan 5's sub-populations were obtained by \cite{Ferraro_2016}  using data acquired with the Hubble Space Telescope’s Advanced Camera for Surveys (HST/ACS) and the Multi-Conjugate Adaptive Optics Demonstrator (MAD) at the ESO
Very Large Telescope (VLT). 
They successfully identified two distinct main-sequence turn-offs (MS-TOs) in the ($K,\,I-K$) CMD. A comparison with isochrones yielded ages of $\sim$12~Gyr for the metal-poor population and 4-5~Gyr for the metal-rich one.  This confirmed that Terzan 5 underwent at least two significant periods of star formation separated by several gigayears.  These ages are consistent with results obtained through star formation history studies, which indicate a major burst of star formation 
12–13 Gyr ago responsible for roughly 70\% of the current stellar mass, followed by additional events at cosmic times around 7 and 4 Gyr \citep{crociati2024}.
Further indication of the presence of the younger sub-population comes from variable star studies. \cite{origlia2019} have identified 
a long period Mira variable whose radial velocity and chemical abundances are consistent with the super-solar population, as expected for a thermally pulsing asymptotic giant branch star arising from an intermediate age component.

The substantial observational evidence has led to growing agreement that Terzan 5 does not fit the conventional definition of a GC, and numerous scenarios for the true nature of this system have been proposed in the literature.
The notion identifying Terzan 5 with an accreted dwarf galaxy, in analogy to the case of $\omega$ Centauri \citep{bekki2003}, is disfavored by the star formation rate (SFR) implied by the observed chemical pattern, since this is incompatible with the SFR of known dwarfs \citep{Tolstoy}. In addition, the reconstructed orbit of the system \citep{Baumgardt2019, baumgardt2021}, with an apocenter of merely 2.8 kpc, strongly argues for an in situ origin \citep{Massari2019,Callingham2022}.
Other mechanisms have been suggested,  including the possibility of rare GC-GC mergers \citep{khopersov2018, mastrobuonobattisti2019}, but these fail to account for the extreme stellar mass of the sub-populations and for the young, metal-rich component, whose properties do not resemble those of any Galactic GC. The alternative scenario involving the interaction between a GC and a giant molecular cloud (hereafter GC+GMC; \citealt{McKenzie_2018, bastianpfeffer}) can in principle trigger a secondary episode of star formation producing stars with features comparable to those of the second observed component. However, these events require fine-tuned conditions and cannot easily explain the presence of a third population or of an extended star formation period (see also discussion in \citealt{dalessandro2022}). Furthermore, the large population of millisecond pulsars  requires 
many type II supernova explosions in a deep potential well to form and retain neutron stars. These conditions arise naturally in scenarios of bulge-like (i.e., high star formation rate) enrichment within a very massive primordial system. 
In turn, these properties, combined with the high collision rate observed in Terzan 5 \citep{Lanzoni2010} can explain the formation of many millisecond pulsars through an efficient recycling of old neutron stars. 
Taken together, these pieces of evidence seem to converge toward the interpretation that Terzan 5 is the remnant of a massive primordial structure that contributed to the early formation of the Galactic bulge—a Bulge Fossil Fragment (BFF; \citealt{Ferraro2021}).
This view links Terzan 5 to theoretical frameworks of bulge formation through the early fragmentation of gas-rich, turbulent disks.  When efficient gas cooling is present, gravitationally unstable disks are expected to break into massive clumps ranging from $10^8$ to $10^9\,M_\odot$ \citep{Immeli2004}. Observational data for such clumps have been obtained across redshifts $z \sim 1$–10: HST has revealed turbulent disks hosting stellar clumps with masses reaching up to $10^9\,M_\odot$ at $z \sim 1-5$ \citep{elmegreen2008, b&f2009, genzel2011}, while JWST has uncovered similar structures at $z \sim 10$ \citep{messa, adamo}. While simulations predict that most clumps migrate inward and merge to form the bulge \citep{ceverino2010}, a small fraction may survive \citep{bournaud2016}. Terzan 5 may represent one such survivor, having avoided destruction by remaining in the bulge's outskirts, which allowed it to retain supernova ejecta and trigger subsequent star formation episodes through dynamic interactions with bulge substructures like the bar.
Terzan 5 does not appear to be a unique case. Liller 1, another massive system ($M \simeq 1.5 \times 10^6\,M_\odot$; \citealt{saracino2015}) traditionally cataloged as GC and located at about $0.8$ kpc from the Galactic center, exhibits remarkably similar properties: multiple stellar populations spanning an age interval from 1–3 Gyr to 12 Gyr \citep{Ferraro2021,dalessandro2022}, a metallicity range of $\sim$1 dex \citep{crociati2023} with an $\alpha$-enhanced old component and solar-scaled young population \citep{deimer2024,2024A&A...690A.139F}, and chemical patterns matching the bulge field \citep{ferraro2025}. 

Improving our understanding of Terzan 5 is thus crucial for providing observational benchmarks to guide future theoretical endeavors to explain the formation of BFFs and galactic bulges. The accurate reconstruction of the formation history of BFFs is crucial also to properly estimate the overall amount of compact objects generated during their evolution, which makes these primordial building blocks of galaxy bulges the most likely efficient factories of gravitational wave sources in spirals (see \citealt{ferraro26a}).

Previous datasets
were limited by severe extinction, differential reddening and crowding, restricting age analyses to small  sub-regions (like the 600 arcsec$^2$ field of view (FoV) used  in \citealt{Ferraro_2016}). 
The James Webb Space Telescope (JWST), with its high performance capabilities at near-infrared wavelengths, presents an exciting new opportunity to study Terzan 5's stellar populations and obtain a deeper, global, and minimally reddening-affected CMD.

As first result of the JWST exploration  of Terzan 5, here we present the deepest and highest-quality CMD ever obtained for this system. The proper motion cleaned and differential reddening corrected CMD provided an unprecedented view of the MS-TO region and allowed us to further investigate  the star formation history of this stellar system revealing that  it is even more complex than previously thought.

This paper is organized as follows: in Section \ref{data} we describe the dataset and the photometric reduction procedures. Section \ref{pm} is devoted to the proper motion computation and the assessment of cluster membership. In Section \ref{reddening} we present the adopted method for differential reddening correction. The isochrone fitting technique and age determination through the MS-TO are presented in Section \ref{age}. Finally, in Section \ref{conclusion} we summarize the main results and discuss their implications.

\begin{figure}[t]
  \centering  \includegraphics[width=0.5\textwidth]{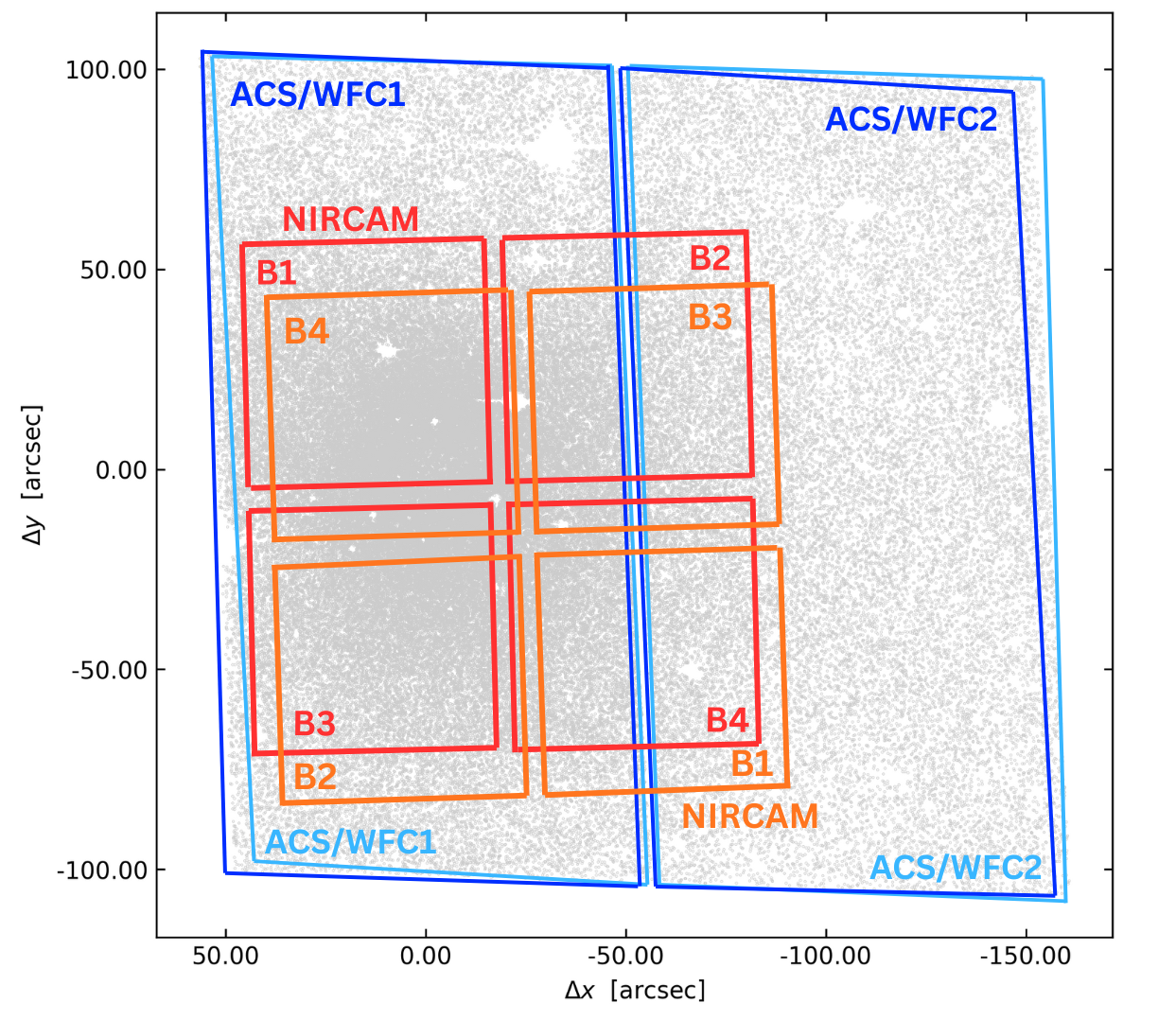}
  \caption{Illustration of the FoV coverage for the dataset of observations of Terzan 5 used in this work. Grey dots represent stars detected in the optical filters. Blue rectangles show the HST/ACS field of view for the first two epochs (SNAP 9799, PI: Rich; GO 12933, PI: Ferraro), while light blue rectangles indicate the third epoch (GO/DD 14061, PI: Ferraro). Red rectangles mark the JWST/NIRCAM module B SW detector boundaries from the first observing cycle (GO 5502, PI: Ferraro, 2024), and orange rectangles show the corresponding fields from the second cycle (GO 5502, PI: Ferraro, 2025). Each of the FoVs shown here corresponds to the pointing of the first exposure in each dataset.  All coordinates are expressed as projected offsets in arcseconds ($\Delta x$, $\Delta y$) relative to the center of Terzan 5 ($\alpha_{\mathrm{J2000}} = 17^{\mathrm{h}}48^{\mathrm{m}}4.85^{\mathrm{s}},\quad
\delta_{\mathrm{J2000}} = -24^\circ46'44.6''$; \citealt{Lanzoni2010}).}
  \label{fig:fov}
\end{figure}

\section{Data analysis}
\label{data}
\subsection{Dataset}
The photometric analysis of Terzan 5 presented in this work relies on a combination of space based near-infrared and optical data, summarized briefly here.
The primary dataset consists of observations obtained with JWST using the Near-Infrared Camera (NIRCam). NIRCam is composed of two identical modules (A and B), each covering a 2.2' $\times$ 2.2' FoV. In each module, four short-wavelength (SW) detectors with a pixel scale of 0.031''/pixel and one long-wavelength (LW) detector with 0.062''/pixel can operate simultaneously, enabling parallel acquisition in multiple filters.
The observations targeting Terzan 5 (Proposal ID: GO 5502, PI: F. Ferraro; see Figure \ref{fig:Terzan 5nircam}) were collected  in two epochs.  
During the first epoch (2024), observations were obtained in both the F115W and F200W SW filters, while the second epoch (2025) consisted exclusively of F115W imaging. The simultaneous LW detector images acquired in filters F277W and F356W were not used for the analysis presented here. We also excluded exposures acquired with 
chip A of NIRCam as they 
sample the outskirts of the system and 
do not overlap with the HST images' FoV. 
Both \texttt{BRIGHT2} and \texttt{RAPID} readout patterns were employed to combine deep and short integrations, ensuring optimal sampling of the magnitude range, from the faintest to the brightest stars.

We also made use of archival optical HST observations obtained with the ACS Wide Field Channel (WFC),  retrieved from the Mikulski Archive for Space Telescopes (MAST). The WFC provides a 0.05"/pixel resolution and a 202" × 202" FoV.
The archival observations span three epochs — 2003 (Proposal ID: SNAP 9799, PI: Rich), 2013 (Proposal ID: GO 12933, PI: Ferraro), and 2015 (Proposal ID: GO/DD 14061, PI: Ferraro) — all in the F606W and F814W filters.
Their main purpose is to provide complementary optical coverage and offer multiple observational epochs over a long temporal baseline to compute proper motions. The illustration of the FoV coverage of the JWST and HST datasets is shown in Figure \ref{fig:fov}. A complete summary of the observational data employed in this work is provided in Table~\ref{tab:observations}.
\begin{table*}[htbp]
\centering
\footnotesize 
\renewcommand{\arraystretch}{1.2}
\setlength{\tabcolsep}{6pt}

\caption{Summary of the JWST/NIRCam and HST/ACS datasets used in this work, including observational epochs, filters, number of exposures and readout modes (if appropriate), and exposure times.
}
\label{tab:observations}

\resizebox{\textwidth}{!}{
\begin{tabular}{ccccccc}
\hline\hline
Instrument & Proposal ID & PI & Date & Filter & Exposures & Exposure Time (s) \\
\hline

\multirow[t]{5}{*}{JWST/NIRCam}
 & GO 5502 & Ferraro & 2024 September 19-20 & F115W & 8 BRIGHT2 & 966.4 \\
 &       &        &                       & F115W & 8 RAPID & 21.47 \\
 &       &        &                       & F200W & 8 BRIGHT2 & 751.6 \\
 &       &        &                       & F200W & 8 RAPID & 21.47 \\
 & GO 5502 & Ferraro & 2025 April 4 & F115W & 8 BRIGHT2 & 966.4 \\
\hline

\multirow[t]{12}{*}{HST/ACS}
 & SNAP 9799 & Rich & 2003 September 17 & F606W & 1 & 340 \\
 &       &      &                   & F814W & 1 & 340 \\
 &       &      &                   & F814W & 2 & 10 \\
 & GO 12933 & Ferraro & 2013 August 18-19 & F606W & 5 & 365 \\
 &       &      &                   & F606W & 1 & 50 \\
 &       &      &                   & F814W & 5 & 365 \\
 &       &      &                   & F814W & 1 & 10 \\
 & GO/DD 14061 & Ferraro & 2015 April 20-21 & F606W & 3 & 397 \\
 &       &      &                   & F606W & 2 & 398 \\
 &       &      &                   & F606W & 1 & 50 \\
 &       &      &                   & F814W & 3 & 371 \\
 &       &      &                   & F814W & 2 & 372 \\
 &       &      &                   & F814W & 1 & 10 \\
 \hline
\end{tabular}
}
\end{table*}
\subsection{Pre-reduction procedures: the JWST calibration pipeline}
The NIR dataset was retrieved from the JWST MAST archive in its uncalibrated (\texttt{uncal}) format. This choice allowed us to fine–tune specific parameters of the JWST Science Calibration Pipeline \citep{bushouse2025} to better suit the requirements of our science case. In particular, we modified some Stage 1 processing steps, where the detector–level corrections are applied to ensure an accurate estimate of the flux recorded in each pixel.
The most significant adjustment consisted in the recovery of information collected during the first segment of each exposure ("Frame 0”). In this frame the level of saturation is modest and limited to the brightest stars thus allowing the measure of stars along the entire red giant branch up to 1 magnitude above the horizontal branch level.
As a result, valuable information is preserved in high–signal regions, such as the most crowded central areas of the stellar system, that would otherwise be lost. Particular care was also devoted to the stage of the pipeline that handles the correction of “snowball” artifacts induced by cosmic-rays.
After extensive testing, the parameter \texttt{expand\_large\_events} was set to \texttt{"False"} to prevent the misidentification of saturated stellar cores as snowballs, which produced spurious artifacts in the crowded field of Terzan 5 (an issue denoted also in \citealt{regan2024snowballs}). 
The subsequent Stage 2 of the pipeline was applied to each exposure without significant modifications.
This performed sky background subtraction, flat-field and flux calibration, and assignment of the World Coordinate System information. For compatibility with \texttt{DAOPHOT}, the flux values that the pipeline expresses in mJy/sr were converted first to DN/s and then multiplied by the single frame time of 10.74 s to obtain raw counts. All images were also corrected using the Pixel Area Map (PAM), which is stored in a dedicated extension of the calibrated data products, to account for variations in projected pixel size across the detector. The resulting calibrated and corrected images constituted the input for the photometric analysis. 
Optical HST/ACS images were downloaded in the \texttt{\_flc} format, meaning calibrated, flat-fielded and CTE-corrected single exposures. They required the application of the PAM correction, which was done using \texttt{stsci.skypac}
and the contained function \texttt{pamutils.pam\_from\_file}.

\subsection{Photometric analysis}
The photometric reduction was performed through a dual approach designed to maximize the potential of the dataset, using two independent software packages.
The main analysis was carried out using the \texttt{DAOPHOT
II} package \citep{stetson1987,stetson2011}.
For each individual image in both the JWST and HST datasets, an optimal PSF model was constructed from bright, isolated stars using the \texttt{PSF} routine. These were subsequently applied to each exposure using \texttt{ALLSTAR} to measure instrumental magnitudes for all sources detected 3$\sigma$ above the local background.
To generate a comprehensive catalog of detected sources, we first produced six independent source lists, one for each filter-instrument combination (F606W/WFC1, F814W/WFC1, F606W/WFC2, F814W/WFC2, F115W/NIRCam, and F200W/NIRCam).  They were produced using \texttt{DAOMASTER}, retaining only sources detected in at least four individual frames within each dataset. Geometric transformations among these sub-catalogs were derived and used to report them all onto a common reference frame corresponding to the ACS/WFC1 detector scale. Using \texttt{DAOMASTER} we cross-identified sources across epochs and instruments to produce one single master list including every star detected in at least one of the available subsets. This master list served as the input for the \texttt{ALLFRAME} algorithm, which automatically detects and fits each of these sources in the individual exposures across all epochs and filters.
At this stage, the joint use of optical and NIR data in the creation of a common master list 
proved to be particularly effective. The greater depth of the JWST observations enables detection of faint sources, providing information on their position and magnitude, which improves their recovery and the treatment of blends in the HST photometry. 
In addition, the spatial resolution and wavelength range of the HST images are better suited for bright stars close to saturation in the NIR data. Overall, the complementary behavior of stars at different wavelengths, combined with the use of a single master list, leads to a more robust PSF fitting and enhances the reliability of the final photometry in all the four filters.
The resulting files containing positions and magnitudes of the fitted sources for each image were then organized again into filter–specific groups, following the same subdivision and requirements adopted for the initial subsets. This approach has already been adopted in the study by \citet{Cadelano_2023} and will be fully described in Pallanca et al. (in prep.). This last step gave the final magnitude values in each filter, homogenized to the reference frame of the first image in each subset. Once these sub–catalogs were produced, they were cross–matched and merged into a single catalog containing all four filters. In this final \texttt{DAOMASTER} run, sources were retained only if they had valid measurements in at least 2 filters.

The JWST images were also independently reduced with the \texttt{jwst1pass} photometric pipeline, an adaptation of the \texttt{hst1pass} code originally developed for HST, and recently extended to JWST  \citep{anderson2022, Libralato_2023}.
This complementary reduction was specifically aimed at recovering reliable fluxes for bright stars approaching the non-linear or partially saturated regime of the NIRCam detectors, where standard PSF-fitting through \texttt{DAOPHOT} becomes unreliable. The \texttt{jwst1pass} pipeline makes use of empirical NIRCam ePSFs \citep{nardiello2023} and applies a perturbation procedure (as described in \citealt{andersonking2006,anderson2008,nardiello2018}) in which isolated bright stars in each image are used to derive specific corrections to the library models. This procedure enables accurate photometry even for stars with mild saturation. Positions and fluxes of stars were thus extracted and multiple images were combined through the iterative use of the \texttt{xym2mat} and \texttt{xym2bar} routines (see \citealt{andersonking2006,anderson2008}). Up to this stage, magnitudes obtained from different SW detectors were kept separate. The resulting four sub-catalogs were then transformed onto a common frame, adopting the B1 chip as reference. 

\begin{figure*}[t]
  \centering
  \includegraphics[width=\textwidth]{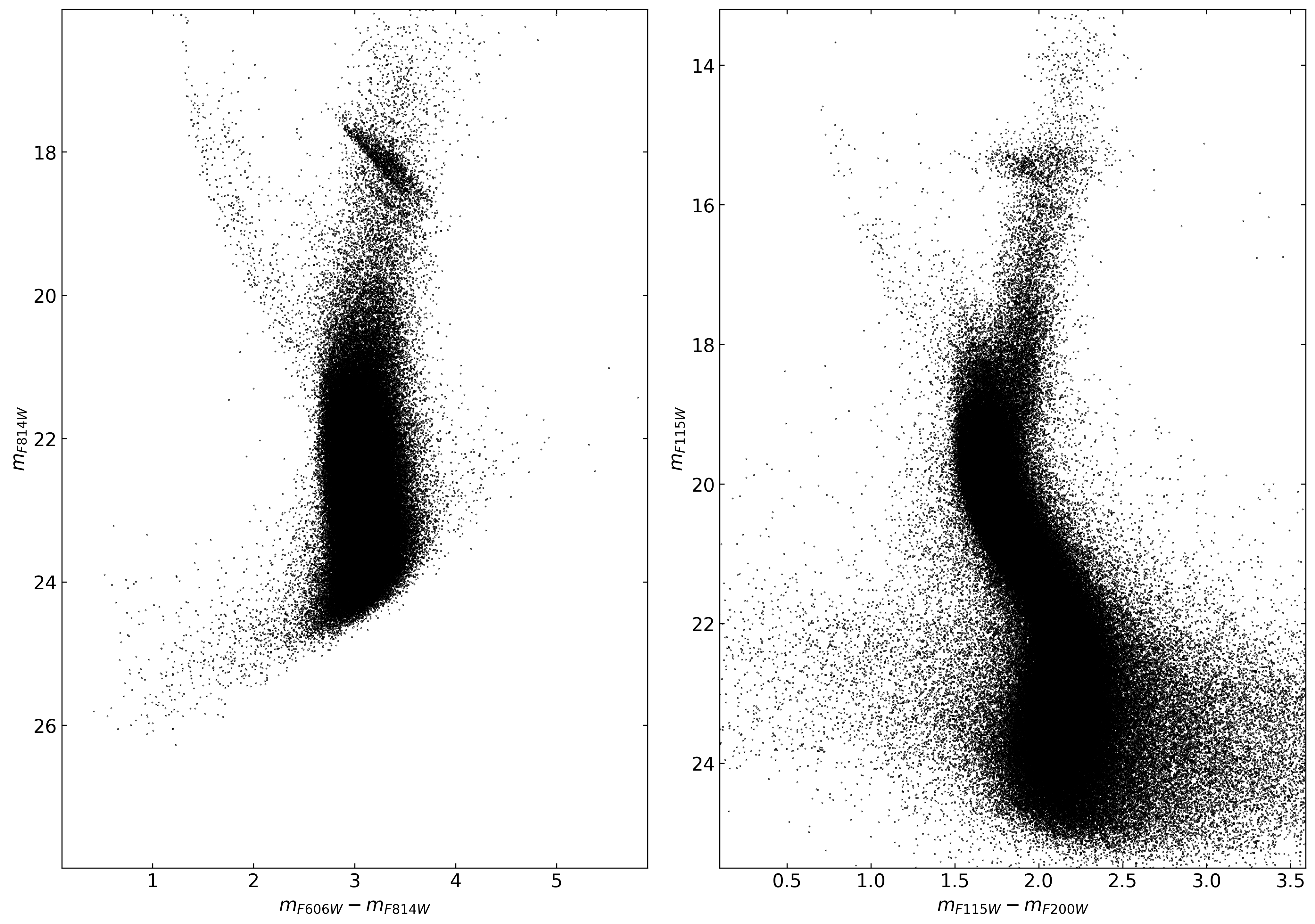}
  \caption{($m_{F814W}, m_{F606W}-m_{F814W}$) and ($m_{F115W}, m_{F115W}-m_{F200W}$) CMDs for Terzan 5 obtained from the joint catalog of HST and JWST datasets. This bipanel visualization highlights the complementary wavelength coverage of optical (left panel) and NIR (right panel) data for studying systems like Terzan 5.}
  \label{fig:cmdtot}
\end{figure*}

\begin{figure*}[htbp]
    \centering
    \includegraphics[width=0.7\textwidth]{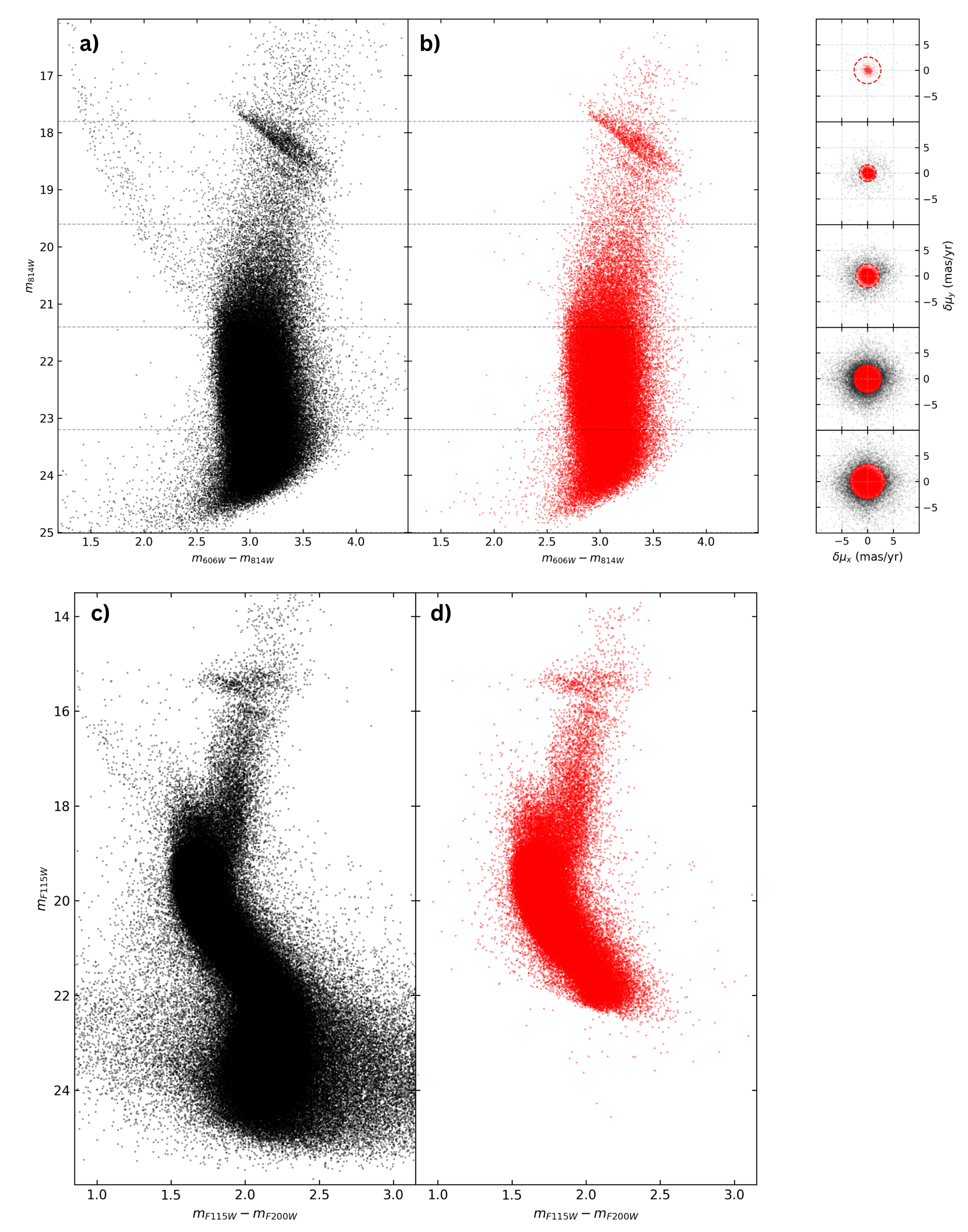}
    \caption{
    Proper–motion selection of Terzan 5 members. Panels a) and b) display the optical CMDs $(m_{F606W}-m_{F814W}, m_{F814W})$ while panels c) and d) show the NIR ones $(m_{F115W}-m_{F200W},m_{F115W})$ . The left panels in each row show in black all the stars that satisfy the photometric–quality criteria, while  panels b) and d) highlight in red only those classified as likely cluster members from the proper–motion analysis. The horizontal dashed lines mark the $m_{F814W}$ magnitude bins adopted in the proper–motion selection. The panel on the upper right shows the corresponding VPDs for each magnitude interval: in each sub–panel, the red dots mark stars retained as members and the black dots indicate rejected (Galactic field) sources.
    The effectiveness of the proper motion membership selection is clearly demonstrated by the rejection of Galactic field MS stars drawing the evident blue plume (bluer and brighter than Terzan 5's MS-TO) in panels a) and c). }  
    \label{fig:pm}
\end{figure*}

\subsection{Photometric and astrometric calibrations}
Both catalogs were calibrated onto the Vegamag photometric system.
For the \texttt{DAOPHOT}-based reductions, instrumental magnitudes were corrected using the Zero Points associated with the reference detector of each filter (Table~\ref{tab:zeropoints}) and adjusted for the corresponding exposure time (340 s for HST and 10.74 s for JWST reference frames).
Aperture corrections were derived as the mean difference between \texttt{PSF} and aperture magnitudes measured with the \texttt{PHOTO} routine, and the missing flux outside the adopted aperture was accounted for using the Encircled Energy corrections (\citealt{jwst_encircled_energy}, \citealt{bohlin2016_encircled_energy}).
Geometric distortion effects were corrected following the prescriptions of \citet{Bellini_2011}.

\begin{figure*}[t]
  \centering
  \includegraphics[width=0.8\textwidth]{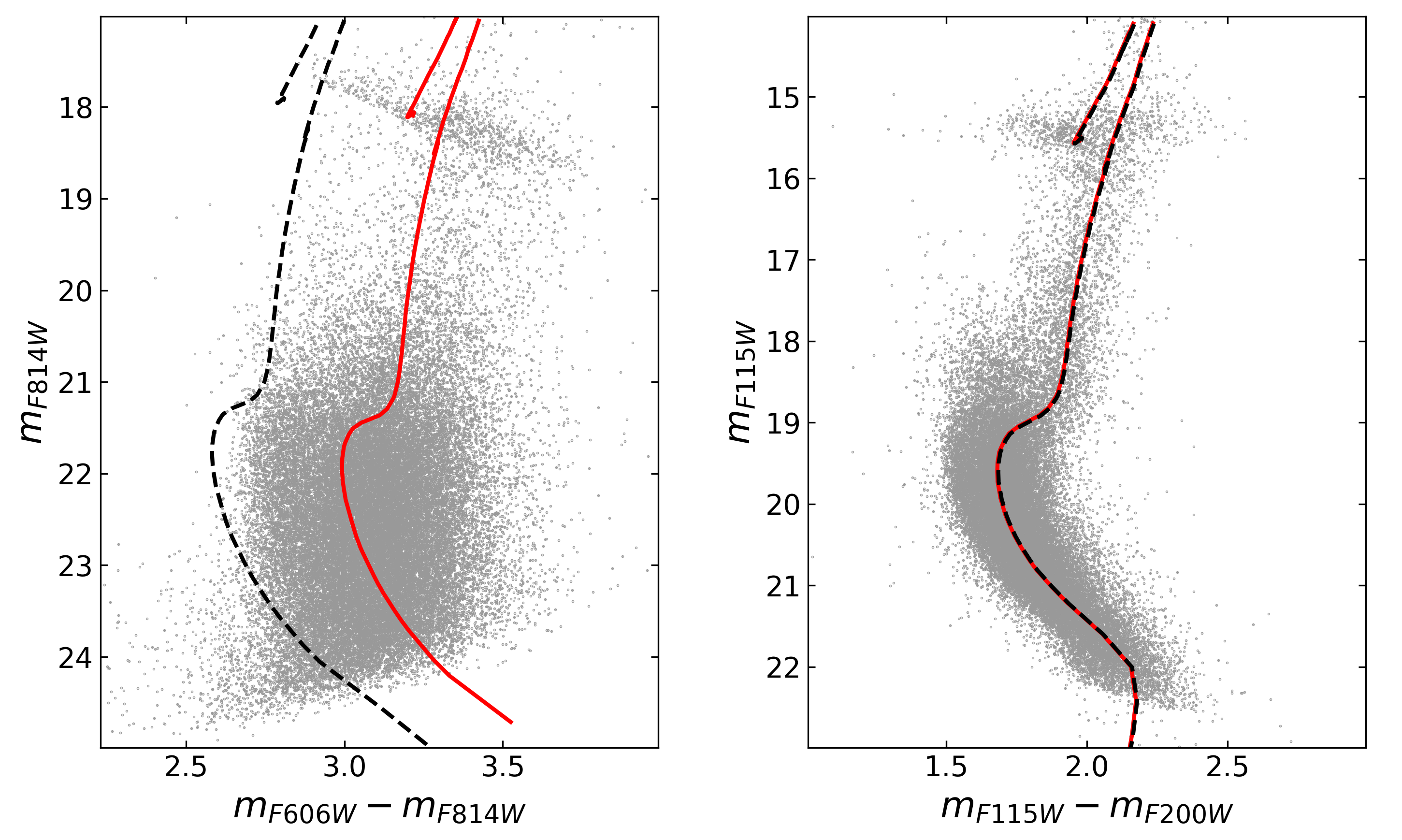}
  \caption{The figure illustrates the effect of different assumptions of the $R_V$ parameter on the isochrone position in the CMDs. The black dashed line is the 12.5 Gyr  isochrone \citep{bressan2012} that best-fits the NIR CMD (right panel) under the assumption of the \citep{odonnell94} extinction law with the standard value $R_V=3.1$, requiring $E(B-V) =2.08$. Clearly, this isochrone is unable to also match the optical CMD (see the black dashed line in the left panel). The red solid line is the same isochrone plotted by assuming the same extinction law, but adopting $R_V=2.5$: it clearly well reproduces simultaneously both the NIR and the optical CMDs, with $E(B-V)=2.86$.
  }
  \label{fig:redlaw}
\end{figure*}
Fluxes measured with \texttt{jwst1pass} were first converted from MJy\,sr$^{-1}$ to DN\,s$^{-1}$, and we then calibrated the four sub–catalogs corresponding to the NIRCam SW detectors using their individual Vegamag zero points. Measurements of stars detected in multiple chips were averaged to yield a single photometric estimate per source.
Aperture and encircled-energy corrections were then computed in the same way as for the \texttt{DAOPHOT} photometry.
Since \texttt{jwst1pass} internally applies geometric-distortion corrections as part of its astrometric solution, no further distortion correction was required.
For both catalogs, we obtained the final astrometric calibration by matching common stars with Gaia DR3 \citep{GaiaDR3}, using the \texttt{CataXcorr} software \citep{Montegriffo1995}.
Finally, the two independently calibrated and astrometrized catalogs were then merged and used jointly for the subsequent analysis.
\begin{table}[htbp]
\centering
\small
\renewcommand{\arraystretch}{1.2}
\setlength{\tabcolsep}{8pt}

\caption{Vegamag zeropoint values for NIRCam and ACS recovered from the NIRCam documentation \cite{jwstzp2024} and the ACS Zeropoint Calculator \cite{acszeropoints}.}
\label{tab:zeropoints}

\begin{tabular}{lcc}
\hline\hline
Filter & Detector & Zero Point \\
\hline
F200W & NIRCam B1 & 25.55 \\
F115W & NIRCam B1 & 25.92 \\
F814W & ACS WFC1/WFC2 & 25.521 \\
F606W & ACS WFC1/WFC2 & 26.420 \\
\hline
\end{tabular}
\end{table}

\begin{figure*}[t]
  \centering
  \includegraphics[width=\textwidth]{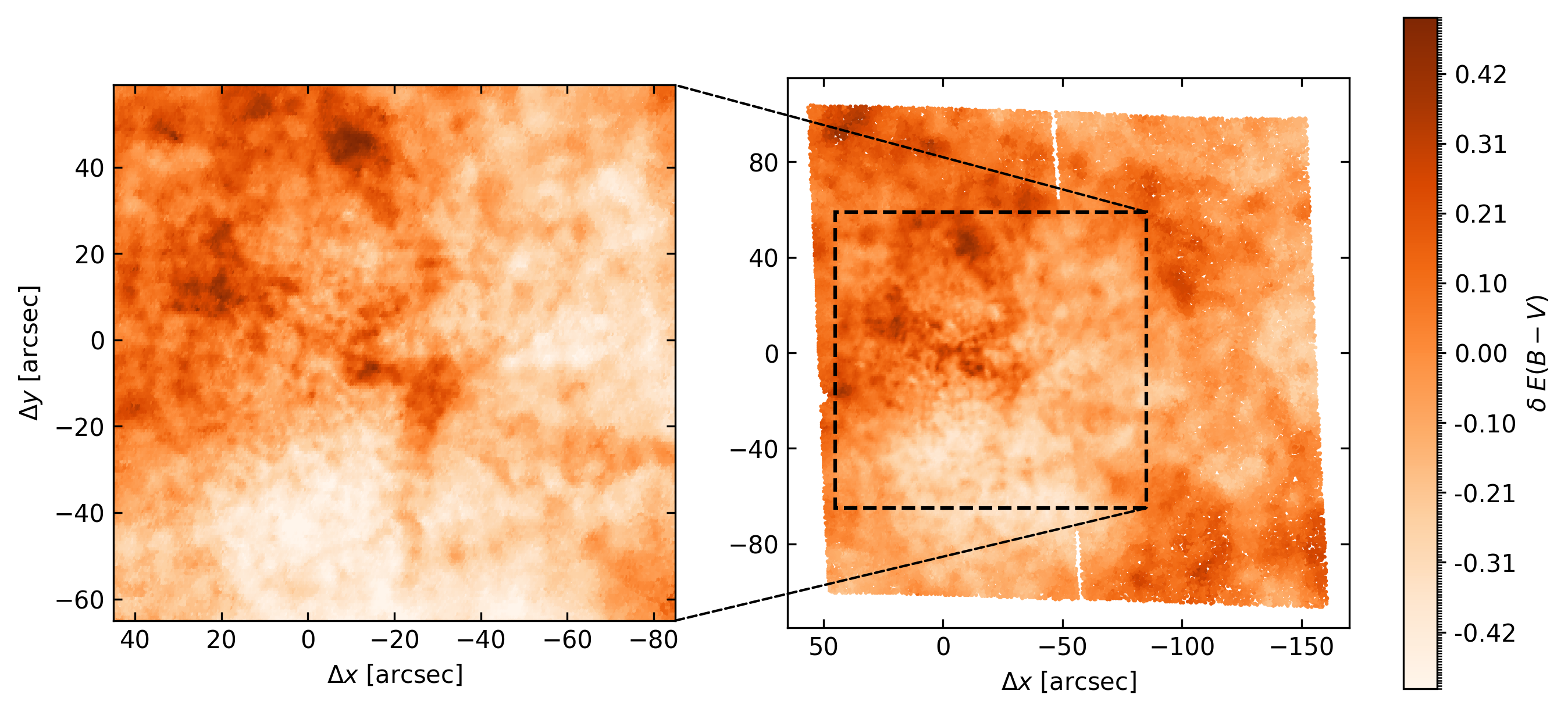}
  \caption{Differential reddening maps derived from JWST/NIRCam (left) and HST/ACS (right) photometry. Both panels show the spatial distribution of $\delta E(B-V)$ across the respective FoVs. The coordinates $\Delta x$ and $\Delta y$ (in arcseconds) indicate the angular offsets from the center of the system. The %shared   [B: perche'~ "shared"?]
  color scale traces variations in $\delta E(B-V)$ with respect to the mean value, highlighting the highly patchy and non–uniform extinction pattern affecting the region.
  }
  \label{fig:mappareddening}
\end{figure*}

\subsection{CMDs of Terzan 5}
Once calibrated, the catalog could be employed for the construction of preliminary CMDs
associated to optical and NIR magnitudes. A comparison of these diagrams is shown in Fig. \ref{fig:cmdtot}. To ensure high–quality photometry in the CMDs, we applied dedicated selection criteria to the HST and JWST datasets. For the \texttt{DAOPHOT} reductions, we retained only sources with \texttt{sharpness},  
\texttt{$\chi^2$}, and magnitude errors consistent with well–fitted point sources. These selections were performed through $\sigma$–clipping, with the acceptable values traced via spline interpolation in successive magnitude bins. For the \texttt{jwst1pass} photometry, we relied on the photometric error and on the quality parameter \texttt{QFIT}, which measures the goodness of the PSF fit: well–measured point sources cluster around \texttt{QFIT} $\approx 0$, and we therefore retained stars with \texttt{QFIT}$<0.3 $ in the final CMDs. An exception was made for stars whose magnitudes were recovered through the dedicated procedure for partially saturated sources, for which the pipeline automatically assigns large \texttt{QFIT} values; in these cases no stringent \texttt{QFIT}-based cut was applied. 

In constructing the CMDs, we made use of both catalogs, with the goal of exploiting the respective advantages of the different reduction methods. After applying the described photometric quality criteria, we compared the source recovery of the \texttt{DAOPHOT} and \texttt{jwst1pass} reductions as a function of magnitude. In the MS-TO region, which is the primary focus of this work and corresponds to $18<m_{\rm F115W}<19.5$, the \texttt{DAOPHOT} catalog contains $\approx 30\%$ more stars than the \texttt{jwst1pass} one. This reflects the advantages of the strategy implemented with \texttt{DAOPHOT} at faint magnitudes in crowded environments, where the joint use of the JWST and HST datasets improves source recovery especially in the central regions. At brighter magnitudes along the red giant branch region (RGB; $m_{\rm F115W}\gtrsim 18$) the number of recovered sources becomes comparable, while approaching $m_{\rm F115W} \simeq 16$, corresponding to the HB level, saturation starts affecting the \texttt{DAOPHOT} measurements. Instead, the \texttt{jwst1pass} reduction remains reliable at these regimes, and we therefore adopt \texttt{jwst1pass} photometry for saturated sources, while using \texttt{DAOPHOT} measurements elsewhere. 

The comparison between the two panels in Fig. \ref{fig:cmdtot} emphasizes the complementary nature of the datasets and
the significant advantage offered by the new NIR observations. JWST confirms its capability of resolving and providing accurate measurements of much fainter magnitudes along
the MS, extending beyond the MS knee.  The JWST/NIRCam diagram is by far the deepest CMD ever obtained for Terzan 5, with the previous
ones \citep{Ferraro_2016} reaching just 1-2 mags below the MS-TO point corresponding to $m_{F115W} \sim 21$.
This newly obtained CMD is thus at least 4 magnitudes deeper than any previous
result. Moreover, the joint reduction with the HST images resulted in an improvement of the optical photometry itself.
\section{Proper motions}
\label{pm}
Terzan 5 lies in a highly crowded region in the bulge direction, which results in significant contamination from stars belonging to the disk and bulge along the line of sight. An effective method to identify genuine Terzan 5 members and decontaminate the CMD from Galactic field interlopers is to rely on proper motion analysis.
The combination with the HST dataset proved to be crucial at this stage, providing a total temporal baseline of more than 20 years.

The first step of the analysis in the adopted procedure 
was the precise determination of stellar centroid positions.
Centroids were extracted from the PSF-fitting results of the photometric reduction, as provided in the output files of the \texttt{ALLFRAME} routine.
JWST coordinates were rectified and transformed onto a common reference system using the routine by \citet{Griggio2023}, while HST data were corrected with the solution by \cite{Bellini_2011}. The images were then grouped by epoch and filter and aligned to the respective reference frames using \texttt{CataXcorr}.
For each detection, the relevant quantities — position, magnitude, error, $\chi^2$, and \texttt{sharpness} — were retained, and a quality selection similar to the one previously explained was applied to select only well-measured stars.
For each epoch, 
the sources satisfying these criteria were used to compute the mean $(x, y)$ positions and associated uncertainties, taken as the standard deviation. When two valid position measurements from different filters were available for the same epoch, their average value was adopted.
Mean stellar positions in detector coordinates were then matched and transformed into a common reference frame, defined by the first-epoch ACS/WFC1 image which covers the central region of the system. 
The alignment was performed with \texttt{CataXcorr} in two steps: an initial transformation limited to first-order terms established a robust preliminary match among catalogs, followed by a higher-order polynomial transformation to refine the solution and maximize the number of matched stars.
Proper motions were derived by performing weighted linear fits of stellar positions as a function of time, using the Modified Julian Dates (MJD) of each observational epoch.
The fits were computed for the $x$ and $y$ directions and weighted by the positional uncertainties, considering only stars with at least two valid measurements across the available epochs. To ensure robustness, only proper motions derived from temporal baselines exceeding three years were retained, since shorter intervals -- such as those involving only the two JWST epochs -- are more affected by positional uncertainties.
The resulting displacements ($\mu_x$, $\mu_y$) were converted into units of mas/yr using the appropriate pixel scale, and associated to stars from the main catalog through cross-matching with \texttt{CataXcorr}.
To separate cluster members from field interlopers, we analyzed the distribution of the total proper motion, $\mu = \sqrt{\delta\mu_x^2 + \delta\mu_y^2}$, in the vector point diagram (VPD) in separate F814W magnitude bins, where $\delta\mu_x$ and $\delta\mu_y$ are the $x$ and $y$ components relative to the mean cluster proper motion. Since proper motions were computed in the cluster's reference frame, the distribution of $\mu$ is expected to be centered around zero for Terzan 5 members. Within each magnitude bin, an iterative $3\sigma$-clipping procedure was applied to the $\mu$ distribution to exclude outliers and retain only stars consistent with the systemic motion of the cluster. The proper motion analysis was implemented through a two–stage approach. After this first iteration, which identified a preliminary sample of likely cluster members, we used 
the latter as a reliable reference frame for a second iteration of the procedure. This last step led to the final proper motion computation and membership selection. 
The results of this selection are illustrated in Figure~\ref{fig:pm}, which shows the VPDs for the adopted magnitude bins.
Because the proper motion measurements rely on the availability of a valid $m_{F814W}$ or $m_{F606W}$ magnitude to guarantee multi–epoch coverage, stars detected only in the NIR do not have associated proper motions. This results in the sharp faint–end truncation observed in the NIR membership–selected CMD of Fig.~\ref{fig:pm}, where sources lacking an optical counterpart cannot be assigned a membership flag.

\begin{figure*}[htbp]
    \centering

    \begin{subfigure}[t]{0.7\textwidth}
        \centering
        \includegraphics[width=\textwidth]{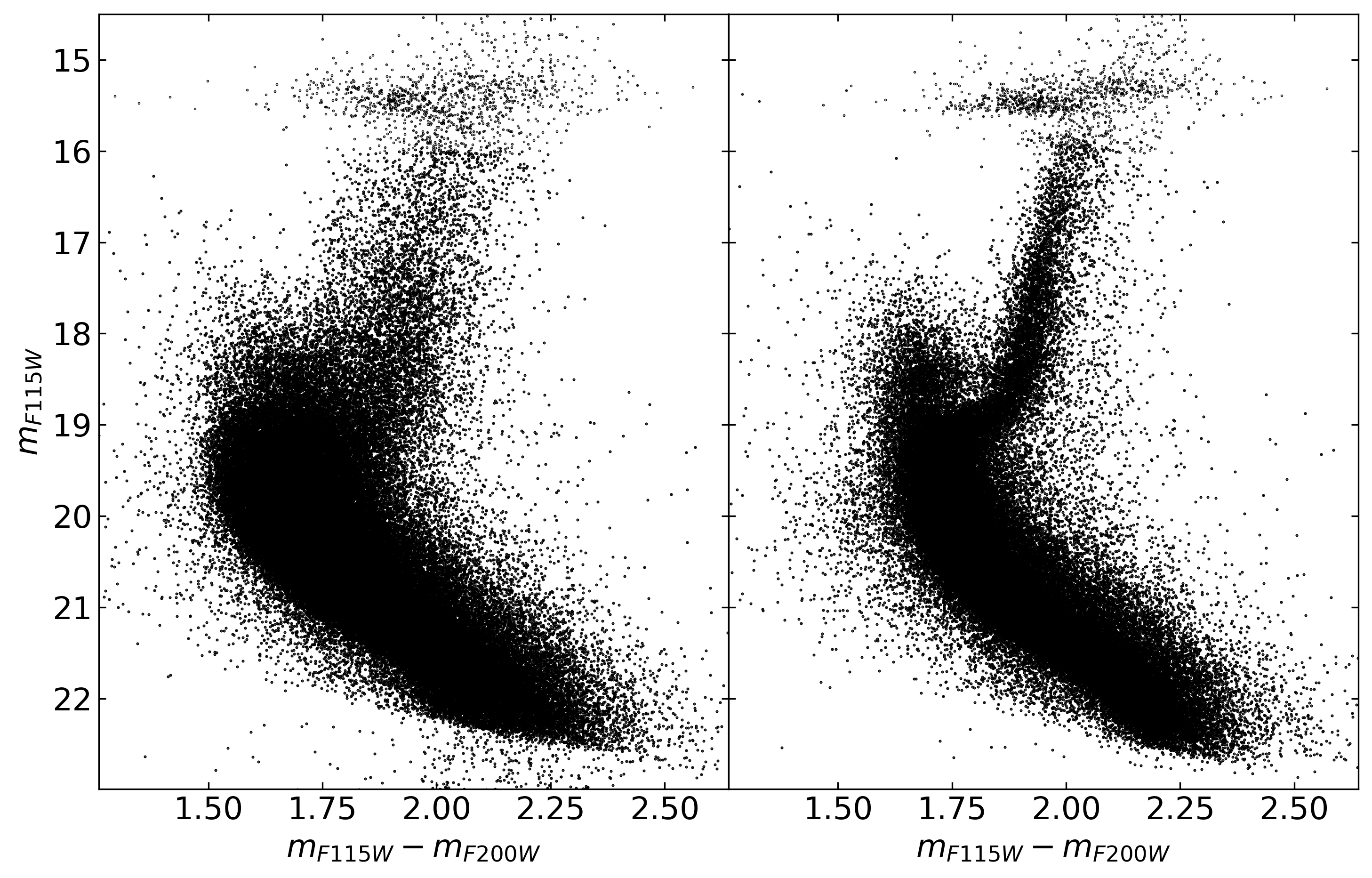}

        \label{fig:postredd_jwst_panel}
    \end{subfigure}

    \vspace{0.8em}

    \begin{subfigure}[t]{0.7\textwidth}
        \centering
        \includegraphics[width=\textwidth]{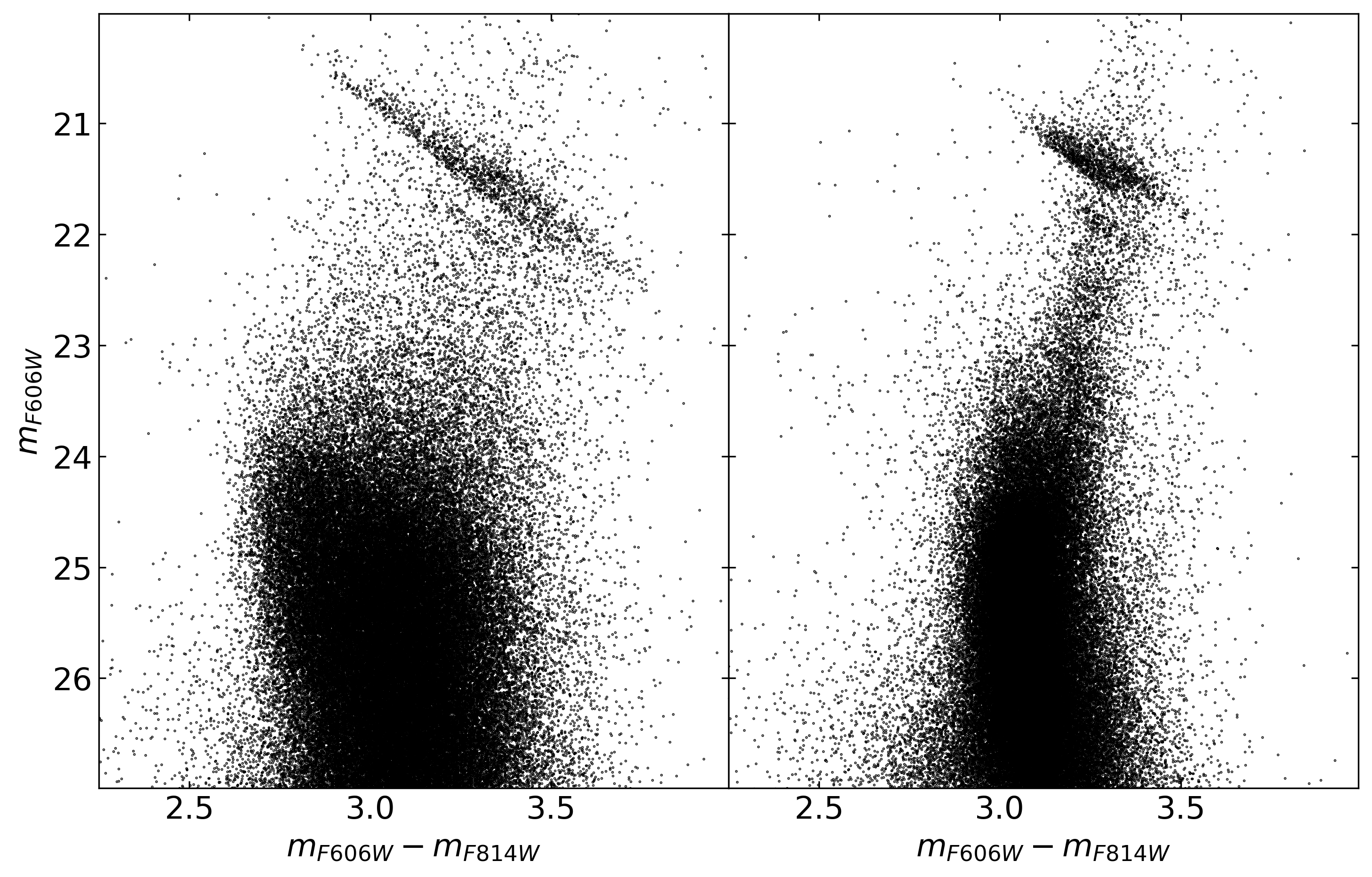}
        \label{fig:postredd_hst_panel}
    \end{subfigure}
    \caption{Effect of  the differential reddening correction on the JWST near-infrared (top panels) 
    and on the HST optical (bottom panels) CMDs. In both cases,  stars have been identified as likely members through proper motion analysis and satisfy strict photometric-quality criteria. The left panels show the original CMDs, while the right panels have the reddening correction applied. We notice how this process significantly reduces color broadening and reveals tighter evolutionary sequences. As expected, the optical CMD is more strongly affected by spatially variable extinction than the NIR CMD.} 
    \label{fig:postredd_both}
\end{figure*}

\begin{figure*}[t]
  \centering
  \includegraphics[width=\textwidth]{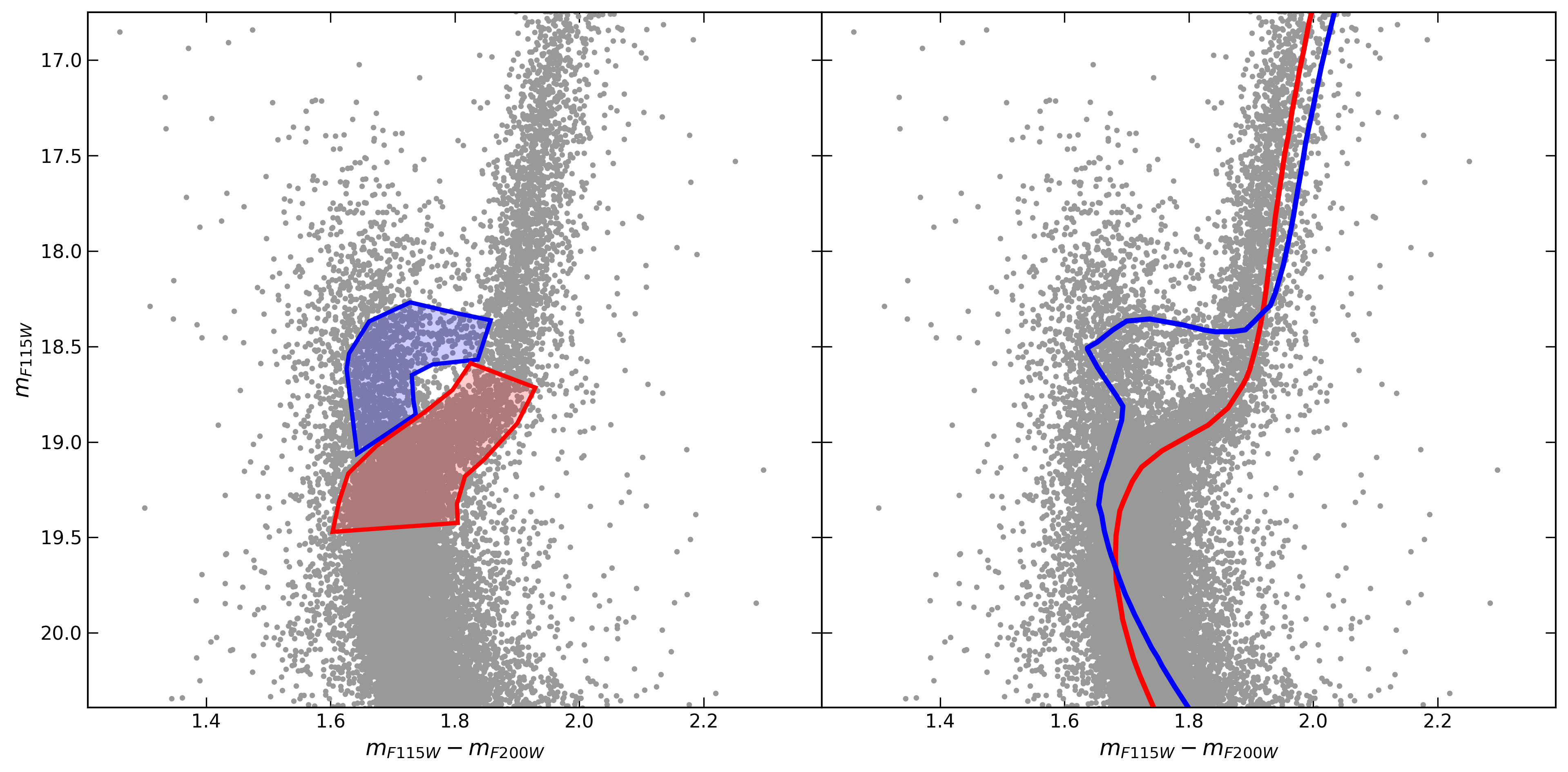}
  \caption{Differential reddening corrected and proper motion selected NIR CMD of Terzan 5. Both panels show stars located between 10" and 40" from the system’s center and satisfying stringent photometric quality criteria to ensure reliable magnitude estimates. Left: the CMD reveals the double MS–TO with unprecedented clarity, definition, and stellar statistics. The superimposed boxes highlight the stars selected along the MS-TO and SGB for the age fitting of the youngest (blue shaded area) and oldest (red) population. Right: the same stars are shown with the best–fitting PARSEC \citep{bressan2012,tang2014,chen2015} isochrones overplotted. The adopted models correspond to ages of 12.5 Gyr (red line) and a 4.7 Gyr (blue line).
  }
  \label{fig:cmdto}
\end{figure*}

\section{Differential reddening}
\label{reddening}
Being located in the bulge,  Terzan 5 is affected by severe and spatially variable extinction along the line of sight \citep{massari2012}. This translates to position-dependent shifts of stars within the CMD because of significant variations of the local $E(B-V)$ with respect to the mean value associated to the system. This produces an apparent broadening of the evolutionary sequences along the direction of the reddening vector. A precise treatment of differential reddening is thus essential for a detailed photometric analysis. Differential reddening in Terzan 5 has already been addressed in previous works from our group using 
different observational datasets 
\citep{massari2012,crociati2024}.
Nevertheless, the availability of the new JWST observations -- providing NIR depth, reduced extinction effects, and a larger spatial coverage compared to the MAD FoV -- calls for an independent determination of the differential reddening correction.
Typically, the wavelength-dependent effect of interstellar extinction is described by laws such as those of \citet{cardelli,fitzpatrick90,odonnell94}, in which the  extinction ratio 
 $R_V=A_V/E(B-V)$ regulates the slope of the attenuation curve. While a "canonical" value of $R_V=3.1$ characterizes the diffuse Galactic interstellar medium along most lines of sight within the Galaxy  \citep{sneden78}, several studies have shown that lower values
(corresponding to steeper extinction laws)
are more appropriate in the Bulge 
direction (e.g., \citealt{Udalski2003,nataf2013,Alonso_Garc_a_2017,pallanca21}). 
Dedicated tests and simultaneous isochrone fitting to the optical and the NIR CMDs show that this is also the case for Terzan 5. Indeed, under the assumption of the standard value ($R_V=3.1$) isochrones provide a reasonable fit to the NIR CMD only, but they fail to reproduce the optical one (see the black lines in Fig. \ref{fig:redlaw}), due to the wavelength dependence of the extinction law. In agreement with previous analyses of star clusters in the Bulge (see \citet{pallanca21, pallanca2021b}), a consistent match across all bands can be achieved only by assuming $R_V=2.5$ (see the red lines in Fig. \ref{fig:redlaw}). By using the \cite{odonnell94} extinction law, this leads to extinction coefficients $R_{F606W}=2.297$, $R_{F814W}=1.433$, $R_{F115W}=0.732$, and $R_{F200W}=0.305$.

By assuming these values, we followed the method by \citet[][see also references therein]{pallanca21}, and we computed the differential reddening correction for Terzan 5.
The procedure first builds the mean ridge line (MRL)  of each CMD by determining the average locus of selected samples of high photometric quality and confirmed member stars along the MS, sub-giant branch (SGB) and RGB. These MRLs provide the fiducial reference against which reddening deviations are measured. For each individual source, we then considered a local sample of nearby reference cluster stars. These stars are required to satisfy quality criteria analogous to those described in the previous section. A grid of trial differential color excess values $\delta E(B-V)$ was applied to the reference stars; for each trial value, 
the CMD positions of the local sample were compared to the MRL, and a distance metric, quantifying residual deviations from the intrinsic MRL, was computed. The optimal 
value of $\delta E(B-V)$ is defined as the 
one that minimizes this distance, and is then converted into magnitude corrections using the adopted extinction coefficients. By construction, this
procedure yields a star-by-star differential reddening estimate, with the highest angular resolution (below $1''$) in the center, which progressively decreases to $\sim 3''-4''$ in the more external regions due to lower local density. 
The resulting two-dimensional reddening maps, constructed independently 
%from 
for the NIRCam and ACS datasets, are shown in Fig. \ref{fig:mappareddening}. Both maps reveal a patchy and spatially variable dust distribution across the FoV, with color excess variations spanning a total range of $\Delta E(B-V) \sim 0.8 $ mag (from approximately -0.4 to +0.4 mag relative to the mean extinction characterizing the adopted MRL). Despite the different wavelength regime and stellar sampling, the maps display consistent morphology in the overlapping area. In the central region of the system, the same large-scale reddening structures and local maxima/minima are recovered by both datasets.
A prominent morphological feature close to the system's center is evident in both maps: an elongated, filamentary structure of extinction $\delta E(B-V) \sim 0.3-0.4$. This consistency strongly validates the robustness of the differential reddening correction method.
The total amplitude of differential reddening  is comparable to that observed in other heavily extincted bulge clusters such as Terzan 6 ($\Delta E(B-V) \sim 0.8 $ mag ; \citealt{Loriga_2025}), Liller 1 ($\Delta E(B-V) \sim 0.5 $ mag; \citealt{pallanca21}), and NGC 6256 ($\Delta E(B-V) \sim 0.5 $ mag; \citealt{cadelano2020}), confirming that Terzan 5 lies in one of these challenging directions for photometric studies in the inner Galactic bulge. 

The obtained values of differential extinction were used to correct the magnitudes of each star, leading to significantly tightened stellar sequences in all CMDs 
(see Fig. \ref{fig:postredd_both}). The proper motion selected and differential reddening corrected CMDs shown in the right panels are those used in all subsequent steps of this analysis, as they best reproduce the intrinsic features of the evolutionary sequences.

\begin{figure*}[t]
  \centering
  \includegraphics[width=\textwidth]{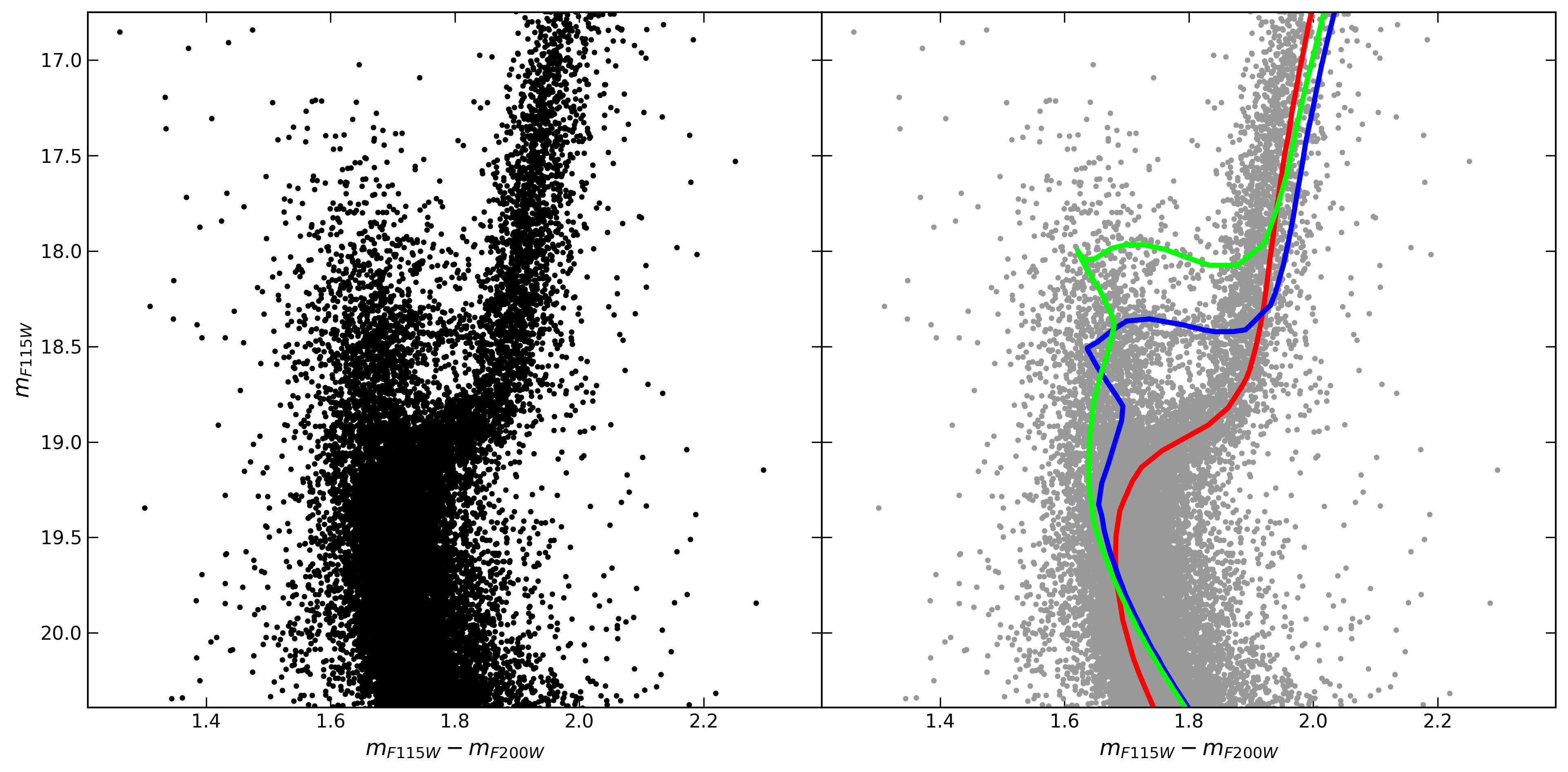}
  \caption{Differential reddening corrected and proper motion selected NIR CMDs of Terzan 5 (in an annular region between 10'' and 40'' from the center). Left: the CMD shows a sequence of stars brighter than the second MS-TO. Right: the same stars are shown in grey, with the best–fitting PARSEC \citep{bressan2012,tang2014,chen2015} isochrones overplotted. The blue and red lines are the same as in Fig. \ref{fig:cmdto}, while the additional component of 3.8 Gyr (Z=0.03) is shown in green. }
  \label{fig:cmdtodoppio}
\end{figure*}

\begin{figure}[t]
  \centering
  \includegraphics[width=0.48\textwidth]{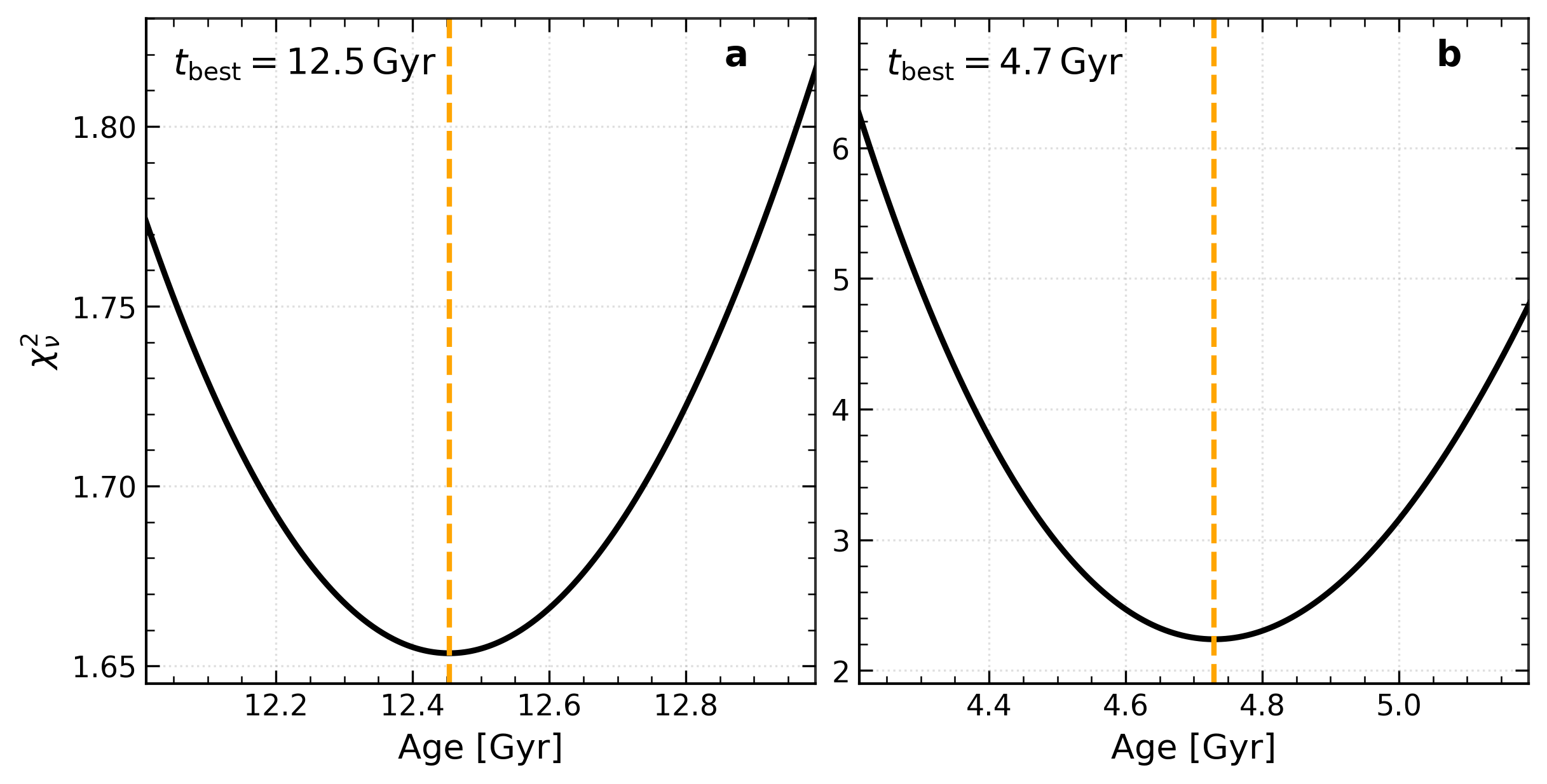}
  \caption{Reduced chi–square ($\chi^2_\nu$) as a function of age values tested for the two main stellar populations identified in Terzan 5, see Fig. \ref{fig:cmdto}. Panel a) refers to the oldest population, panel b) to the younger one. The black solid curves show the parabolic fits to the $\chi^2_\nu$ distributions,  
  while the orange dashed vertical lines mark the best-fitting ages (corresponding to the $\chi^2_\nu$ minimum).}
  \label{fig:chiquadri}
\end{figure}

\section{Age determination}
\label{age}
\subsection{Constraining the color excess and distance modulus}
\label{ebv_mu0}
Isochrone fitting in a complex system hosting numerous distinct populations is intrinsically challenging. For this reason, we attempted to simplify the problem by developing a two-step procedure. In the first step, we aimed to constrain global parameters characterizing all the populations, such as the mean extinction ($E(B-V)$) and the distance modulus ($\mu_0$). To this aim, we
selected 
isochrones from the PARSEC \citep{bressan2012,tang2014,chen2015} stellar evolution library, spanning ages and metallicities consistent with those reported in the literature for the oldest (12-13 Gyr) sub-solar population (see \citealt{ferraro2009}; \citealt{Ferraro_2016}; \citealt{Massari2014}, \citealt{origlia2025}). The isochrones were then shifted in magnitude and color by adopting the \cite{odonnell94} extinction law with $R_V=2.5$, and assuming several trial pairs of values for $E(B-V)$ and $\mu_0$. These quantities were treated as free parameters together with age and metallicity. The models were compared to the data through a $\chi^2$ metric in the  
CMD, computed as:
\begin{equation}
\chi^2 \;=\; \sum_i \min_j \Bigg[
\begin{pmatrix}
\Delta C_{ij} & \Delta J_{ij}
\end{pmatrix}
\begin{pmatrix}
\sigma_{K,i}^{-2} & -\sigma_{K,i}^{-2}\\[2pt]
-\sigma_{K,i}^{-2} & \dfrac{\sigma_{J,i}^2+\sigma_{K,i}^2}{\sigma_{J,i}^2\,\sigma_{K,i}^2}
\end{pmatrix}
\begin{pmatrix}
\Delta C_{ij}\\[2pt]
\Delta J_{ij}
\end{pmatrix}
\Bigg],
\end{equation}
where $\Delta C_{ij}$ and $\Delta J_{ij}$ are the color and F115W magnitude differences between the $i$-th star and the $j$-th point along the isochrone, and $\sigma_{J,i}^2,\sigma_{K,i}^2$ are the photometric error variances in the F115W and F200W magnitudes for that star, respectively. The fitting procedure was carried out simultaneously on optical, NIR, and hybrid CMDs, considering the whole extent of the evolutionary sequences. This approach exploits the distinct reddening and luminosity sensitivity of each dataset, thus reducing degeneracy between age, metallicity, reddening, and distance. Exploring the parameter space across a dense grid of color excess and distance modulus values yielded the overall best–fitting solution, corresponding to 
$E(B-V)=2.86$ and 
$\mu_0=14.15$. These values were subsequently fixed and adopted for all stellar populations in the second step of the analysis, where the ages of the individual sub–populations were independently determined using 
the MS-TO.
Although the resulting extinction is slightly different from some literature estimates (see \citealt{valenti2010}, \citealt{massari2012}), the discrepancy is expected given the adoption of a different extinction law. The derived distance modulus, on the other hand, is fully consistent with recent the independent estimate for Terzan 5  ($d = 6.78 \pm 0.22$ kpc, corresponding to $\mu_0=14.15$) obtained with the latest Gaia DR3 data \citep{baumgardt2021}.
\subsection{A new age estimate for Terzan 5's sub-populations}
Having fixed the extinction and distance modulus, we proceeded to derive the ages of the stellar populations hosted in Terzan 5. The analysis was carried out 
on the NIR CMD that has the best quality at the MS-TO level. The presence of two clearly distinct MS-TOs can be appreciated with unprecedented detail in the left panels of Fig. \ref{fig:cmdto} and Fig. \ref{fig:cmdtodoppio}.
To reduce blend contamination, we considered only stars in an annular region located between 10" and 40" from the system’s center. In addition, stringent cuts on the \texttt{sharpness} parameter were applied to further suppress residual contamination from unresolved blends \citep{stetson1987}.
For quantitative age estimates, we adopted the isochrone fitting technique, 
searching for the isochrones that best reproduce the observed  MS-TO and SGB region of the two main sub-populations of Terzan 5, as drawn by member stars included within appropriate selection boxes in the differential reddening corrected NIR CMD (see the left panel of Fig. \ref{fig:cmdto}). 
We first extracted isochrones of different ages (with a 0.1 Gyr step) from the PARSEC \citep{bressan2012} database, by adopting metallicities consistent with the spectroscopic measurements for the two main components of the system ([Fe/H]$\approx-0.3$ and [Fe/H]$\approx+0.3$; \citealt{Massari2014, origlia2011}). We therefore assumed $Z=0.0077$, $Y=0.26$ for the sub-solar population, and $Z=0.03$, $Y=0.29$ for the super-solar one
(with the adopted values of the helium abundance following the standard enrichment scenario; \citealp{girardi2002}).
Each model was then shifted according to the derived best-fit values of $E(B-V)$ and $\mu_0$ (Section \ref{ebv_mu0}), and we used the $\chi^2$ test to identify the isochrones that best fit the data.  

The resulting $\chi^2$ vs age profiles 
(Fig. \ref{fig:chiquadri}) exhibit well–defined minima, yielding two distinct best–fitting solutions: an old population with an age of $12.5 \pm 0.5$ Gyr and a younger population with age $4.7 \pm 0.5$ Gyr. The formal statistical uncertainties from the parabolic fit to the $\chi^2$ profile are respectively $\pm 0.02$ Gyr and $\pm 0.01$ Gyr. The final quoted uncertainties conservatively take into account this statistical error, together with systematics mainly due to intrinsic uncertainties in stellar models and degeneracies between $E(B-V)$ and metallicity. The best-fit isochrones are shown  in the right panel of Fig. \ref{fig:cmdto}. The adopted solutions reproduce well the observed morphology of both the MS-TO and the SGB, supporting the reliability of the derived ages. 
We verified that small variations in the boundaries of the adopted selection boxes do not significantly affect the resulting ages.
We also repeated the fitting procedure with BaSTI isochrones \citep{Pietrinferni2024} of matching metallicity. As expected because of the different underlying parameterizations and physical assumptions of the model sets, BaSTI requires slightly shifted parameters ($E(B-V) = 2.92$ and $\mu_0 = 14.05 $). Nevertheless, the resulting best–fit ages (12.7 and 5.1 Gyr) are consistent with the PARSEC values and fall within the conservative total age uncertainty of $\pm 0.5$ Gyr.
\subsection{Evidence for an additional younger component}
\begin{figure}[t]
  \centering
  \includegraphics[width=0.45\textwidth]{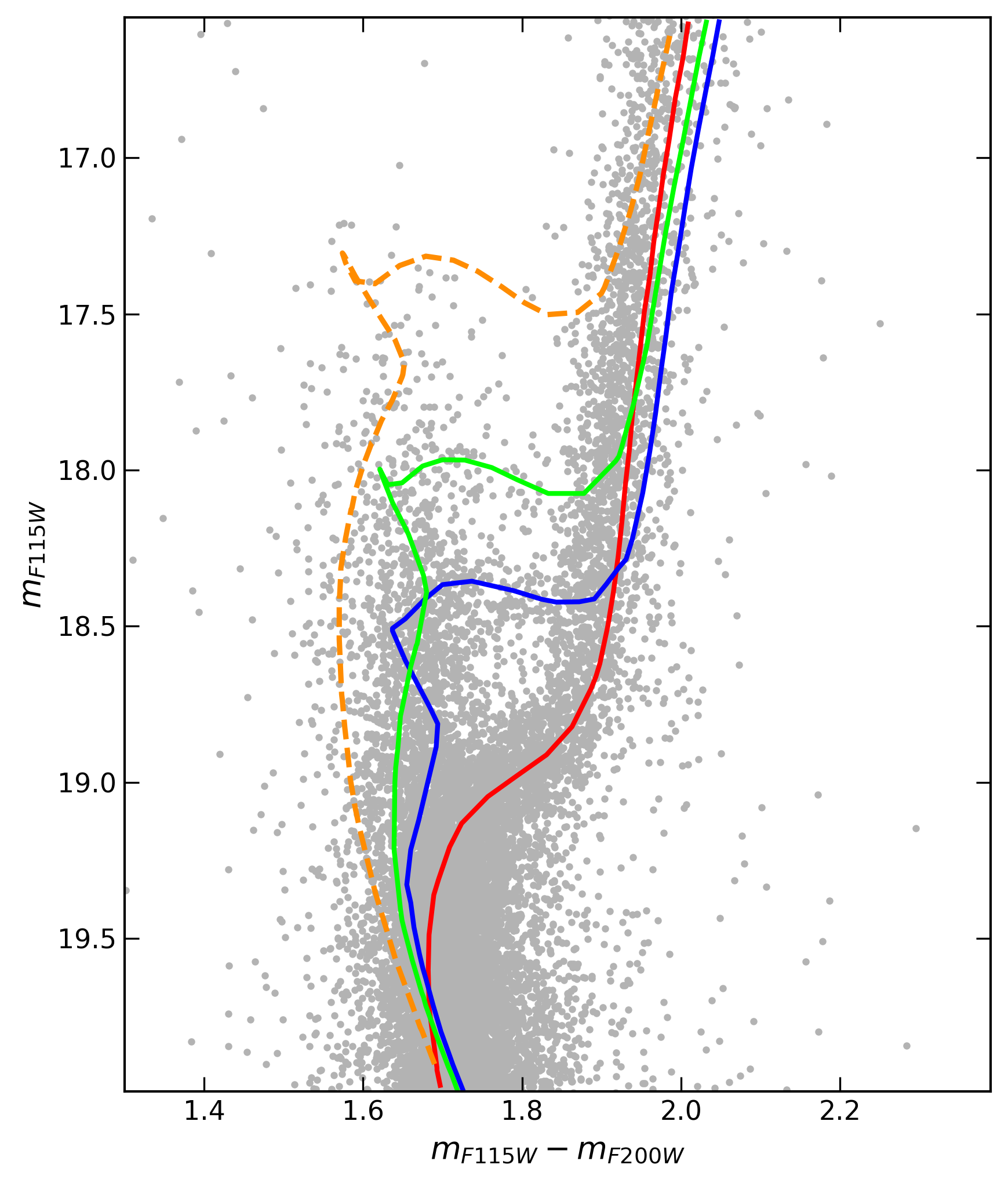}
  \caption{ NIR CMD of Terzan 5 as in Fig. \ref{fig:cmdtodoppio}, with the solid lines tracing the 12.5, 4.7, and 3.8 Gyr old sub-populations discussed in text (red, blue, and green color, respectively), and the dashed orange line corresponding to an isochrone of just the 2.5 Gyr, which is needed to encompass the entire extension of the blue plume.}
  \label{fig:cmdpiuiso}
\end{figure}

One of the most intriguing characteristics of the new JWST CMD 
is the presence of a prominent “blue plume” of stars brighter than the 4.7 Gyr MS-TO (see the sources above the blue line in the right panel of Fig. \ref{fig:cmdto}). This could be the signature of a prolonged, recent star formation. Indeed, the careful inspection of the CMD zoomed in the MS-TO region reveals an alignment of stars in this region of the CMD, possibly tracing the presence of an additional SGB. 
To estimate the age of this possible additional component, we applied to these stars the same $\chi^2$ statistics discussed above. In the absence of spectroscopic 
measures, we tentatively adopted a metallicity comparable to that of the 4.7 Gyr component. 
The right panel of 
Fig. \ref{fig:cmdtodoppio} shows 
the best-fit isochrone (green line), with an age of 3.8 Gyr. We notice,  however, that the blue plume extends to even brighter magnitudes (reaching $m_{\rm F115W} \approx 17.4$) corresponding to stellar masses for which no detectable SGB is expected due to the short evolutionary timescales. In Fig. \ref{fig:cmdpiuiso} we show that the extension in luminosity of the blue plume could, in principle, trace ages as young as $\approx$ 2.5 Gyr. 
The presence of a blue plume naturally raises the question of a possible contribution from blue straggler stars (BSSs; see, e.g., \citealt{ferraro1997,ferraro18,dalessandro2008}), which indeed populate a similar region of the CMD. However, their expected number is generally modest even in the most massive stellar systems (e.g., $\sim 300$ BSSs are observed in $\omega$Centauri; see \citealt{ferraro2006}). To estimate the number of BSSs expected in Terzan 5, we took advantage of the recent results of \citet{ferraro26b}, who found that the BSS  demography strongly depends on the environment. More specifically, they found that the BSS frequency (defined as the number of BSSs per units of $10^4 L_\odot$ sampled luminosity) significantly decreases for increasing central density ($\rho_0$) and central surface brightness ($\mu_0$) of the parent cluster.
Hence, by adopting the relations listed in Supplementary Table 1 of \citet{ferraro26b} and the values of $\rho_0$ and $\mu_0$ quoted in \citet{Lanzoni2010}, we estimate that approximately 3 BSSs are expected for each $10^4 L_\odot$ of sampled luminosity in Terzan 5. By considering that the WFC3 dataset samples approximately $5\times 10^5 L_\odot$, we expect that approximately 150 BSSs contaminate the observed blue plume population, which comprises $\sim 1600$ stars in total. Moreover, BSS formation is expected to occur in sporadic events, rather than tracing a coherent activity. Hence, the clear presence of well-defined MS-TOs and well-populated SGBs provides strong evidence of  "standard" star formation bursts occurred in the recent past of this stellar system, rather than random BSS formation events. We therefore conclude that, while some BSSs are likely present in this region of the CMD, their contribution is negligible and cannot account for the observed extent of the blue plume.

\subsection{Spatial distributions of the  stellar populations}
\begin{figure}[t]
  \centering
  \includegraphics[width=0.5\textwidth]{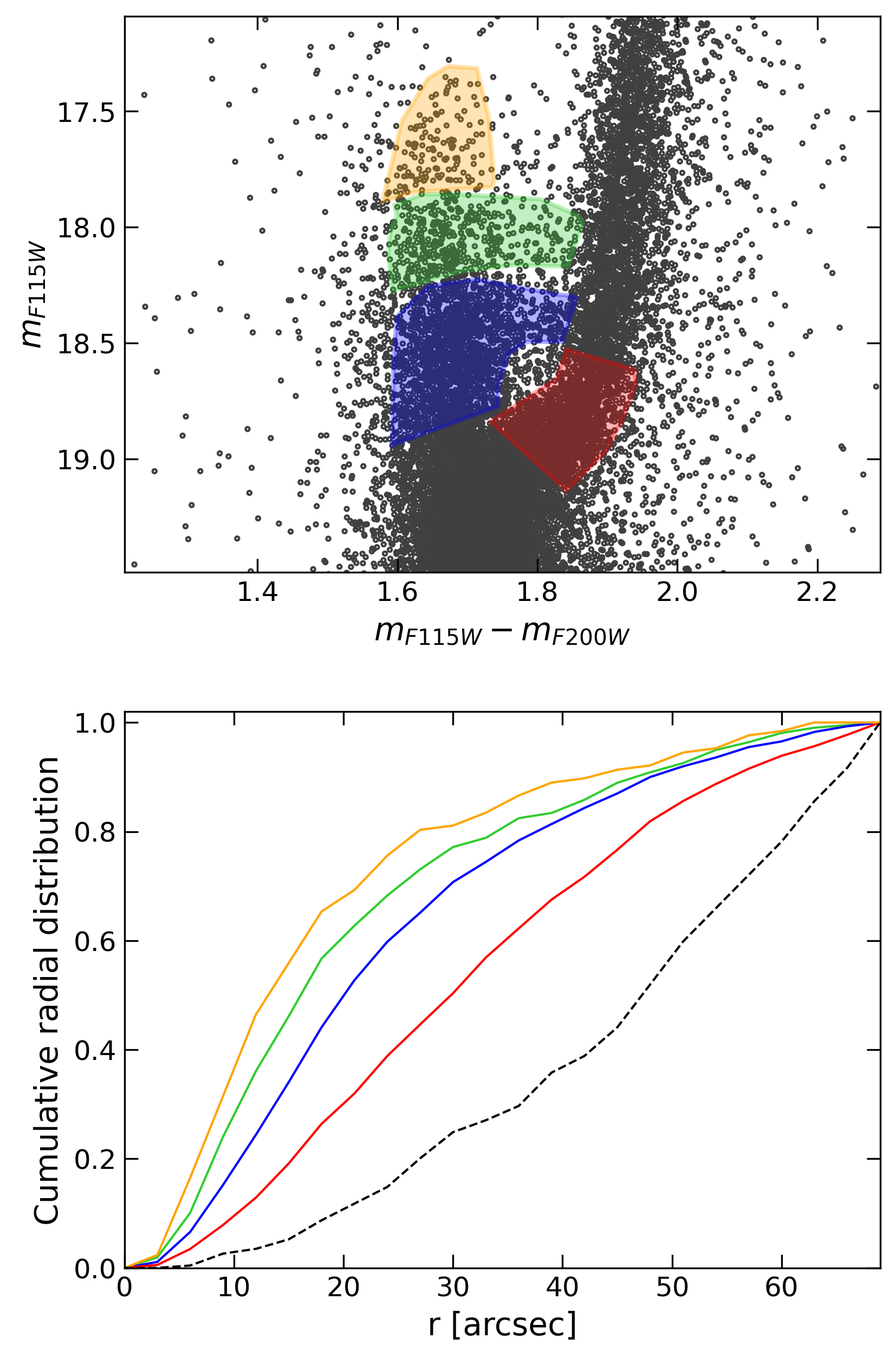}
  \caption{Upper panel: JWST CMD with stars in the 0"-70" annular region from the center zoomed in the MS-TO region. The selection boxes are the ones adopted to extract the sub-populations of different ages, each highlighted with a different color: red, blue, green and yellow colors for the components with and age of 12.5, 4.7, 3.8 Gyr, and younger than 3.8 Gyr, respectively.  
  Lower panel: cumulative radial distributions of the selected components (solid lines with the same color-code as in the upper panel). A sample of Galactic field stars (black dashed line) is also shown as reference.}
  \label{fig:radialdist}
\end{figure}

To further investigate the nature of these stellar populations, we selected three corresponding samples in the MS-TO region of the CMD 
%as shown in 
(see the upper panel of Fig. \ref{fig:radialdist}), and analyzed their cumulative radial distributions. Because the considered populations span a small magnitude range ($\delta m_{F115W}\approx 1.5$ mag) the comparison of their cumulative radial distributions is expected to be minimally affected by completeness differences. As shown in the 
bottom panel of the same
figure, 
the younger the population is,
the more centrally concentrated it appears:
the stars belonging to the
newly identified 3.8 Gyr old component (marked in green) are the most centrally segregated, followed by those belonging to the 4.7 Gyr population (in blue), and then
by the oldest, 12.5 Gyr old, population (in red). In turn, all the three components are more segregated toward the center of Terzan 5 than Galactic field stars (marked in
black), as expected for populations that do belong to the system. Moreover, the stars selected along the brightest part of the blue plume (in yellow), exhibit an even higher degree of central concentration.

\section{Summary and conclusions}
\label{conclusion}
In this work we have presented a JWST exploration of 
the MS-TO region of the heavily extincted stellar system Terzan~5 orbiting the Galactic bulge, with the aim of determining the ages of its multi-iron stellar sub-populations with unprecedented precision. Our main findings can be summarized as follows:

\begin{enumerate}
    \item We obtained the deepest NIR CMD of Terzan 5, reaching more than four magnitudes fainter than previous studies.
    \item 
    The joint reduction of JWST and HST images enabled refinements in the PSF fitting, which led to significant improvements in the photometric quality in all four available filters. The inclusion of HST images was also crucial to compute  relative proper motions that allowed us to distinguish member stars from field interlopers. 
    \item Simultaneous 
    isochrone fitting in the optical and NIR CMDs provided 
    us with the appropriate extinction law in the direction of Terzan 5 ($R_V=2.5$). This extinction law was then adopted in the construction of a high-resolution differential reddening map of the system. 
   \item After cleaning for proper motions and correcting for differential reddening, the CMD finally revealed a complex structure of the MS-TO region of Terzan 5, with two main sub-populations (with ages of $12.5\pm 0.5$ Gyr, and $4.7\pm 0.5$ Gyr), a possible third component with an age of 3.8 Gyr, and an extended blue plume reaching $m_{F115W}\sim 17.4$, which is likely populated by stars as young as 2.5 Gyr.
    \item The radial distributions of these populations display increasing central concentration with decreasing age.
\end{enumerate}

As a first crucial result, this work provides the definitive confirmation of the presence of two main distinct MS–TOs in Terzan 5 (Fig. \ref{fig:cmdto}). They are now resolved with a level of precision never reached before. The newly derived ages are fully consistent with those reported in \citet{Ferraro_2016}, which have been obtained across different photometric bands, thereby strengthening and consolidating the conclusion that this stellar system hosts a sub-solar metallicity component with an age as old as ($12.5\pm 0.5$) Gyr, together with a much younger metal-rich population of ($4.7\pm 0.5$) Gyr.
In addition, the JWST exploration has unveiled the existence of an extended blue plume populated by Terzan 5 member stars. Their distribution in the CMD reaches almost one magnitude above the MS-TO of the 4.7 Gyr old component, and is suggestive of a prolonged activity of star formation extending up to just 2.5 Gyr ago (Fig. \ref{fig:cmdpiuiso}). The evidence of a sort of "bridge" at $m_{F115W}\approx18$ suggests that, on top of this low-intensity activity, the system experienced at least an additional burst of star formation $\sim 3.8$ Gyr ago (Fig. \ref{fig:cmdtodoppio}). On the other hand, the different sub-populations are systematically more centrally concentrated for decreasing age (Fig. \ref{fig:radialdist}), a feature that naturally fits into a self-enrichment scenario.

These findings provide new observational constraints that help narrowing the range of viable interpretations of the true nature of Terzan 5. In particular, the clear identification of a prolonged period of star formation, possibly characterized by multiple bursts, strongly disfavors  the GC+GMC interaction scenario \citep{McKenzie_2018}. Together with the different radial segregation observed for the different age components, these features are instead suggestive of a self-enrichment process, where subsequent generations of stars form in the system’s central regions from gas retained and processed by previous generations.
In this context, the existence of stellar populations younger than the 4.7 Gyr component suggests that stars with metallicity higher than [Fe/H]$=+0.3$ dex are present in the system. Indeed the survey by \citet{Massari2014} showed some preliminary hints of the possible presence of an extreme, super metal-rich population reaching [Fe/H]=$+0.8$. Moreover, the self-enrichment history of a $\sim 4 \times 10^7 M_\odot$ system that underwent major episodes of star formation has been proven to well reproduce the ages and metallicities of the two main sub-components (see \citealp{romano2023}). Finally, these new findings are in agreement with the star formation history of Terzan 5 reconstructed in \cite{crociati2024}, who found indications of a continuous activity with a few individual bursts. 

Overall, the present investigation makes the case of Terzan 5 even closer to that of the other BFF discovered so far (Liller 1), which also presents an extended blue plume in the CMD (see \citealp{Ferraro2021}), and is characterized by a continuous, low-intensity star formation activity extending approximately over the last 10 Gyr, with some major bursts at specific epochs (see \citealp{dalessandro2022}). These might be the signatures of dynamical interactions with bulge sub-structures (e.g., the bar), therefore bringing new information about the past evolution of the heart of our galaxy. 
Future spectroscopic follow-up will be essential to confirm the nature of the youngest sub-populations hosted in Terzan 5 and further reconstruct the star formation and evolutionary history of this extraordinary system. In a broader context, the transformative role of JWST in the exploration of highly extincted and crowding-limited environments finally opens the way for analogous analyses of other bulge GCs, which might disclose similar peculiarities and therefore reveal to be other remnants of massive structures that contributed to the formation of the Milky Way bulge.

\begin{acknowledgements}
This work is part of the project {\it "GENESIS - Searching for the primordial structures of the Universe in the heart of the Galaxy"} (Advanced Grant FIS-2024-02056, PI:Ferraro), funded by the Italian MUR through the {\it Fondo Italiano per la Scienza (FIS)} call.

Davide Massari acknowledges financial support from PRIN-MIUR-22: CHRONOS: adjusting the clock(s) to unveil the CHRONO-chemo-dynamical Structure of the Galaxy” (PI: S. Cassisi). 
\end{acknowledgements}

\bibliographystyle{aa}
\bibliography{bibliography}

\end{document}